\documentclass[fleqn,usenatbib]{mnras}
\usepackage{newtxtext,newtxmath}

\usepackage{float}
\usepackage[T1]{fontenc}
\DeclareRobustCommand{\VAN}[3]{#2}
\let\VANthebibliography\thebibliography
\def\thebibliography{\DeclareRobustCommand{\VAN}[3]{##3}\VANthebibliography}

\usepackage{amsmath,amssymb,graphicx}
\usepackage{aas_macros}
\usepackage[dvipsnames]{xcolor}
\newcommand{\lcdm}{\mathrm{\Lambda CDM}}
\newcommand{\npipe}{\texttt{NPIPE}}
\newcommand{\camspec}{\texttt{CamSpec}}
\newcommand{\quickpol}{\texttt{QuickPol}}
\newcommand{\thetamc}{\theta_\mathrm{MC}}
\newcommand{\thetastar}{\theta_*}
\newcommand{\ns}{n_\mathrm{s}}
\newcommand{\As}{A_\mathrm{s}}
\newcommand{\ombh}{\Omega_\mathrm{b} h^2}
\newcommand{\omch}{\Omega_\mathrm{c} h^2}
\newcommand{\taure}{\tau_\mathrm{reio}}
\newcommand{\omk}{\Omega_K}
\newcommand{\neff}{N_\mathrm{eff}}
\newcommand{\mnu}{m_\nu}
\newcommand{\planck}{\textit{Planck}}
\newcommand{\plik}{\texttt{Plik}}
\newcommand{\hillipop}{\texttt{HiLLiPoP}}
\newcommand{\rchisq}{\hat \chi^2}
\newcommand{\cl}{C_\ell}

\newcommand{\muK}{\mu \mathrm{K}}
\newcommand{\muKsq}{\mu \mathrm{K}^2}

\newcommand{\eg}{EG21}
\newcommand{\plkiii}{PPL18}

\newcommand{\npp}{NP20}

\newcommand{\pmapsiii}{HFI18}

\providecommand{\sorthelp}[1]{}
\def\GHz{\ifmmode $\,GHz$\else \,GHz\fi}

\title[Cosmological parameters from Planck PR4]{CMB power spectra and cosmological parameters from \planck{} PR4 with CamSpec}
\author[Rosenberg, Gratton \& Efstathiou]{
Erik Rosenberg,\thanks{E-mail: \href{mailto:er510@cam.ac.uk}{er510@cam.ac.uk}}
Steven Gratton and
George Efstathiou
\\
Institute of Astronomy and Kavli Institute for Cosmology Cambridge, Madingley Road, Cambridge, CB3 0HA, UK
}

\date{Accepted XXX. Received YYY; in original form ZZZ}
\pubyear{2022}

\begin{document}
\label{firstpage}
\pagerange{\pageref{firstpage}--\pageref{lastpage}}
\maketitle

\begin{abstract}
    We present angular power spectra and cosmological parameter constraints derived from the \planck{} PR4 (\npipe{}) maps of the Cosmic Microwave Background. \npipe{}, released by the Planck Collaboration in 2020, is a new processing pipeline for producing calibrated frequency maps from \planck{} data. We have created new versions of the \camspec{} likelihood using these maps and applied them to constrain $\lcdm{}$ and single-parameter extensions. We find excellent consistency between \npipe{} and the \planck{} 2018 maps at the parameter level, showing that the \planck{} cosmology is robust to substantial changes in the mapmaking. The lower noise of \npipe{} leads to \textasciitilde{}10\% tighter constraints, and we see both smaller error bars and a shift toward the $\lcdm{}$ values for beyond-$\lcdm$ parameters including $\omk$ and $A_L$.

\end{abstract}


\section{Introduction}
\label{sec:introduction}

The \planck{} survey of the anisotropies of the cosmic microwave background (CMB) has been invaluable to cosmology. It has tightly constrained the $\lcdm{}$ model parameters, ruled out many plausible extensions, and as an important legacy dataset is now frequently combined with ground-based CMB data and other cosmological probes. \planck{} analysis has evolved significantly in this time, from the initial 2013 data release \citep{planck2013-p01} to the addition of polarization in 2015 (PR2) \citep{planck2014-a01}, and substantial improvement of systematic corrections in 2018 (PR3) \citep{planck2016-l01, planck2016-l06}. Further work on the data processing and mapmaking led to the 2019 publication of \texttt{SRoll2} \citep{delouis2019}, an extension of 2018's \texttt{SRoll} destriping algorithm \citep{planck2014-a10}, and later the \npipe{} release in 2020 \citep{planck2020-LVII}. \npipe{} is a new and independent pipeline to produce frequency maps from the time-ordered data, with substantial differences in detector calibration and systematic corrections compared to previous releases. In this paper we present a new version of the \camspec{} likelihood, developed in previous \planck{} likelihood papers \citep{planck2013-p08, planck2014-a13, planck2016-l05} and described in detail in \citealt{eg21}. We use \npipe{} 100, 143, and 217\GHz{} maps to calculate pseudo-power spectra and covariances and so build a high-$\ell$ \npipe{} likelihood. The 353 and 545\GHz{} channels are also used as galactic dust templates. We compare the power spectra and parameter constraints obtained to those found with the 2018 maps, allowing us to evaluate the robustness of \planck{} cosmology to quite substantial changes in the data processing.

We begin by summarizing the \npipe{} mapmaking pipeline (Sec.\ \ref{sec:npipeintro}), the data and masks we use (Sec.\ \ref{sec:data}, \ref{sec:masks}), and the dust-cleaning methodology (\ref{sec:dust}). We show details of the \npipe{} temperature-to-polarization (TP) leakage and effective polarization efficiencies in Sec.\ \ref{sec:calib}. Our noise estimation method and \npipe{} noise spectra are presented in Secs.\ \ref{sec:noise} and \ref{sec:corrnoise}, with the latter focusing on noise correlations between data splits. \npipe{} power spectra are shown in Sec.\ \ref{sec:powspec}, along with tests of internal consistency. We present parameter constraints from the new \npipe{} likelihoods in Sec.\ \ref{sec:likelihood} and explore extensions to $\lcdm{}$ in Sec.\ \ref{sec:extensions}. We compare to an alternative \npipe{} likelihood as well as to ACT and SPT in Sec.\ \ref{sec:comparisons} and conclude in Sec.\ \ref{sec:conclusions}.
For conciseness we refer to the \planck{} 2018 high frequency instrument (HFI) mapmaking and likelihood papers \citep{planck2016-l03, planck2016-l05} as \pmapsiii{} and \plkiii{}. Similarly, we call the \camspec{} paper \citep{eg21} \eg{} and the \npipe{} paper \citep{planck2020-LVII} \npp{}.

\section{Introduction to \npipe{}}
\label{sec:npipeintro}

The \npipe{} pipeline is described in detail in \npp{}, including a comparison of the processing and resulting maps to PR3, a description of the simulations, and an estimate of the optical depth to reionization $\tau$. Here we summarize the main aspects of the pipeline relevant to this work, focusing on HFI.  

\subsection{\npipe{} preprocessing}
\label{sec:npipepreprocessing}
The first stage of the \npipe{} pipeline is preprocessing, performed on single detectors and mostly on single pointing periods. This is analogous to the construction of cleaned time-ordered data (TOD) in previous \planck{} releases (\citet{planck2014-a08} and \citet[sec. 2.1]{planck2016-l03}).

The first processing stage involves estimation of the sky signal from raw bolometer data, for use in subsequent steps. \planck{} scanned a given circle on the sky \textasciitilde 39-65 times before moving on to the next. Within one such pointing period the data from each revolution may be binned together to estimate the sky signal on the circle. These estimates, referred to as (\texttt{HEALPix}\footnote{\url{http://healpix.sourceforge.net}}) ring maps, were used extensively in PR3 and earlier releases. \npipe{} instead prefers a `global signal estimate' in all cases except glitch removal. The global signal estimate involves sampling full-mission maps from an earlier iteration of \npipe{} into the time domain, then adding an estimate of the orbital dipole and convolving by the bolometer transfer function. This signal estimate is intended to be less noisy than that derived from rings, and also allows for the use of `repointing' data, taken between the pointing periods.

Apart from the signal estimation, the most significant differences between \npipe{} preprocessing and that done previously are in the handling of 4K lines and glitches.
As previously \citep{planck2014-a08}, 4K lines are fitted for each pointing period with sine and cosine templates at selected harmonic frequencies of the 4K cooler electronics in the time domain. \npipe{} fits for six more harmonics seen in some detectors. For the six strongest (highest S/N) lines \npipe{} also chooses to use a shorter fitting period than the full pointing period so as to reduce line residuals due to the time-varying amplitudes and phases of the lines.
When removing glitches -- large residuals in the data due primarily to cosmic ray hits -- \npipe{} uses a phase-binned estimate of the signal for each pointing period, similar to the PR3 rings. However \npipe{} employs wider bins, and uses as the signal estimate a polynomial fit to the data in a given bin plus its two neighbors instead of a simple mean. The additional smoothing leads to a less noisy signal estimate, reducing half-ring correlations and allowing the glitch threshold to be increased, reducing the number of flagged glitches by about half. Finally, the bolometer time response deconvolution step has been modified to minimize the effects of the gaps left by glitch removal on the science data, reducing small-scale noise and half-ring correlations.

\subsection{\npipe{} reprocessing and destriping}
For the reprocessing stage \npipe{} uses a generalized destriper \citep{keihanen2004, keihanen2005, keihanen2010} similar to \texttt{SRoll} \citep{planck2014-a10} to perform calibration and systematic template fitting. Writing the data model $ \mathbf{d} = \mathsf{P}\mathbf{m} + \mathsf{F} \mathbf{a} + \mathbf{n}$ we have time-ordered data $\mathbf{d}$, pointing matrix $\mathsf{P}$, sky map $\mathbf{m}$, and a white noise term $\mathbf{n}$. $\mathsf{F}$ is a matrix of time-domain templates representing systematic and astrophysical effects to be removed, with $\mathbf{a}$ the amplitude of each template. \npipe{} solves for template amplitudes $\mathbf{a}$, using them to make a template-cleaned TOD $\mathbf{d}_\mathrm{clean} = \mathbf{d} - \mathsf{F}\mathbf{a}$. Since the template for gain fluctuations is derived from the estimated sky map, the cleaned TOD is used to update estimates of $\mathsf{F}$ and $\mathbf{m}$ and the process iterated until the template amplitudes converge to zero.

While using a similar principle to \pmapsiii{}, \npipe{} differs significantly in the templates $\mathsf{F}$ used. 
One major change is in the imposition of a prior on the polarization maps during calibration. This is done to resolve a degeneracy created between gain and polarization residuals due to the \planck{} scanning strategy. In effect fluctuations are allowed in the gain which are compensated by spurious polarization signal. To address this problem \npipe{} follows the methods of LFI \citep{planck2016-l02} and uses the 30, 217, and 353\GHz{} polarization maps as templates in $\mathsf{F}$. This allows the gains to be solved on an effectively unpolarized sky map, removing the degeneracy. Because the polarization templates are foreground dominated, the destriper will try to fit the CMB signal with the other templates in $\mathsf{F}$. This suppresses the CMB polarization signal and leads to a large-scale polarization transfer function in \npipe{}.

Another significant change in \npipe{} is the use of much shorter baselines for low-frequency noise offsets. $\mathsf{F}$ models $1/f$ noise as a constant offset for each baseline period; \pmapsiii{} used a single such offset per ring, while in \npipe{}  each corresponds to $1^\circ$ on the sky. A single pointing period would therefore include thousands of these shorter baselines. \npp{} finds that this method better captures low-frequency instrumental noise, leading to lower noise at $\ell > 20$.

Finally, \npipe{} introduces two time domain polarization templates to measure corrections to polarization efficiency and polarization angle, and also adds a template to better account for systematics from the analog-to-digital converter nonlinearity (ADCNL). In addition to allowing for ADCNL-induced effective gain fluctuations as in \texttt{SRoll}, \npipe{} includes a `distortion' template corresponding to effective gains that change as a function of input signal level. This is done to further reduce temperature-to-polarization leakage from the ADCNL.

\subsection{Other differences between \npipe{} and PR3}

In making maps from independent subsets of the data from which to calculate cross-spectra, \npipe{} chooses to split into sets of detectors, referred to as A and B, rather than the half-mission split used in \eg{} and \plkiii{}. The \npipe{} detector sets are similar to the detector-set splits explored extensively in \eg{} but also group one or two spiderweb bolometers with each set of polarization sensitive bolometers. 
More importantly however the \npipe{} reprocessing steps -- destriping and systematic template fitting -- are performed independently on each set. This is in contrast to PR3 in which the full-mission data set at each frequency was fit for systematics (gains and offsets, bandpass corrections, transfer function, far side lobes) together and used for all map products. Indeed \texttt{SRoll} explicitly minimizes the $\chi^2$ difference in signal between one bolometer and the average of all bolometers in a frequency band  when solving for the template amplitudes; this can introduce correlations in the subset maps. This suggests \npipe{} detector sets should be more independent than those of PR3, as we test in Sec.\ \ref{sec:corrnoise}.

Finally, we note that \npipe{} reports an 8\% increase in integration time from including repointing maneuver data, taken between the stable pointing periods as the satellite reoriented from ring to ring. This is reflected in lower noise levels as discussed in Sec.\ \ref{sec:noise}, although it does not account for the entire difference.

\section{Methods and power spectra}
\label{sec:methods}

To create an \npipe{} likelihood we begin by following the methods of \eg{}, generating analogues of the 12.5HMcl likelihood described there. As in previous \camspec{} releases we use the pseudo-$\cl$ method \citep{hivon02,kogut03} in which power spectra are computed on masked skies and a coupling matrix is applied to deconvolve the effect of the mask. In this section we discuss foreground cleaning, calibrations and systematic corrections subsequently applied to the deconvolved spectra. We also introduce the noise modeling and investigate correlated noise in \npipe{}.

\subsection{Data}
\label{sec:data}
We use as our starting point \planck{} HFI maps at 100, 143, 217, 353, and 545\GHz{}. For our PR3-based analysis we use the 2018 half-mission maps with associated pixel covariances from the \textit{Planck} Legacy Archive (PLA)\footnote{\url{https://pla.esac.esa.int/pla}}. As in \eg{} and previous releases the main likelihoods presented in this paper use data-split differences to estimate the noise power, and we compare to results from end-to-end simulations. For PR3 we estimate noise with the FFP10 simulations, available on the PLA, and half-mission odd-even ring differences.

For \npipe{} we use the A/B detector-set maps at the same frequencies, available at NERSC\footnote{National Energy Research Scientific Computing Center, see \url{https://crd.lbl.gov/divisions/scidata/c3/c3-research/cosmic-microwave-background/cmb-data-analysis-at-nersc/}}, with the associated \verb|wcov_mcscaled| files for pixel covariances.  We use the \npipe{} simulations and half-ring A/B maps from NERSC for noise estimation.
 We subtract the PR2 solar dipole before using the frequency maps, noting that imperfections in this subtraction should not affect our high-$\ell$ likelihoods. The \npipe{} low-$\ell$ EE transfer function is neglected as we do not use $\ell < 30$. 

To deconvolve the instrument beams we employ transfer functions $W_\ell$ calculated with \quickpol{} \citep{quickpol} and provided with each data release.
\subsection{Sky Masks}
\label{sec:masks}
The sky masks used here are identical to those used in \eg{} and described in \citet{planck2014-a13}, up to corrections for `missing pixels' which we address momentarily. As previously we use diffuse foreground masks to remove strong galactic dust emission and also mask regions with significant emission from point sources, CO, and extended objects by applying additional `point source' masks. In all our likelihoods we use a diffuse mask that retains 80\% sky area before apodization (mask80) at all frequencies in temperature and polarization. Each frequency has a different point source mask in temperature, and the 143\GHz{} point source mask is used for all frequencies in polarization. 

\planck{} half-mission maps contain a number of pixels (from 2,500 for 100\GHz{} half-mission 1 to 115,000 in 217\GHz{} half-mission 2) marked as missing due to poor determination of the polarization in those pixels. \npipe{} maps do not contain any such missing pixels, owing to the inclusion of data from repointing maneuvers as well as changes to the glitch detection. Accordingly no missing pixels are masked in our \npipe{} likelihoods.

\subsection{Dust Cleaning}
\label{sec:dust}
The likelihoods presented here all include `dust cleaning', using higher frequency \planck{} maps to remove the effects of galactic dust at the power spectrum level as in Eq.\ \ref{eq:spectraclean}:
\begin{equation}
  \begin{split}
    C^{X_{\nu_1}Y_{\nu_2}\mathrm{clean}} &= \left(1+\alpha^{X_{\nu_1}}\right)\left(1+\alpha^{Y_{\nu_2}}\right)C^{X_{\nu_1}Y_{\nu_2}} \\
    & - (1 + \alpha^{X_{\nu_1}})\alpha^{Y_{\nu_2}}C^{X_{\nu_1}Y_{\nu_T}} \\
    & - (1 + \alpha^{Y_{\nu_2}})\alpha^{X_{\nu_1}}C^{Y_{\nu_2}X_{\nu_T}}  + \alpha^{X_{\nu_1}}\alpha^{Y_{\nu_2}}C^{X_{\nu_T}Y_{\nu_T}}.
   \label{eq:spectraclean}
   \end{split}
\end{equation}
The $\alpha^{X_\nu}$ are cleaning coefficients at frequency $\nu$; $C^{XY}$ is a mask-deconvolved, beam-corrected power spectrum where $X, Y \in \{T, E, B\}$; $\nu_T$ is the template frequency, here 353\GHz{} for TE and EE spectra and 545\GHz{} for TT spectra.

The cleaning coefficients $\alpha^{X_\nu}$ are calculated by minimizing the function $\sum_{\ell_{\mathrm{min}}}^{\ell_{\mathrm{max}}} C_\ell^{X_{\nu}X_{\nu}\mathrm{clean}}$.
This minimization gives a biased coefficient $\alpha'$ and so we instead use the corrected value $\alpha = \alpha' / (1 + \alpha')$ as the cleaning coefficient.
In temperature we minimize over the $\ell$-range $100 \leq \ell \leq 500$, where galactic dust dominates over the Cosmic Infrared Background (CIB). For polarization spectra we use $30 \leq \ell \leq 300$. To limit the impact of noise from the 353\GHz{} maps in polarization, we fit the dust contribution with a power law and at $\ell > 150$ subtract this fit from the uncleaned spectra rather than using the cleaned spectra.

As demonstrated in \eg{} this method is very effective at removing dust emission in both temperature and polarization. This allows all TE and EE spectra to be coadded, and no foreground nuisance parameters are included for polarization in the likelihood. For temperature we include nuisance parameters describing power law residuals for each frequency combination, accounting for power from residual CIB, point sources, and other foregrounds. At 100\GHz{} large-scale foregrounds are not removed as accurately as at 143\GHz{}, as 100\GHz{} maps contain CO emission and non-negligible synchrotron emission that are difficult to capture accurately in even a more complicated foreground model. 100\GHz{} is also noisier at high multipoles and the primary CMB temperature fluctuations are already excellently determined by 143\GHz{} and 217\GHz{}. Therefore we omit 100\GHz{} from the temperature likelihoods.

\subsection{Calibration, polarization efficiencies, and TP leakage}
\label{sec:calib}
\subsubsection{Temperature Calibration}
We apply multiplicative calibration factors to each temperature spectrum entering the likelihood, interpreting these differences as small transfer function errors due to mismodeling of the beam outside the main beam. The calibration factor $c_{{\nu_1}{\nu_2}}$ for a temperature spectrum $D_\ell^{\nu_1\nu_2}$ is calculated by minimizing the $\chi^2$ of Eq.\ \ref{eq:calmin}:
\begin{equation}  
  \chi^2 = \sum_k \frac{\left(c_{{\nu_1}{\nu_2}}D_k^{{\nu_1}{\nu_2}} - D_k^{143\times 143}\right)^2}{\sigma^2_{{\nu_1}{\nu_2}}}.
\label{eq:calmin}
\end{equation}
The sum is over bandpowers $k$ in the multipole range $50 \leq \ell \leq 500$, and $\sigma$ is the scatter of those bandpowers over the fitted multipole range. The $143 \times 143$ spectrum is subtracted so as to eliminate cosmic variance. In order to remove foreground effects before minimization we also apply a conservative galactic mask (mask30), the 217\GHz{} point source mask, and 545\GHz{} cleaning to all frequencies. The resulting temperature calibration factors $c_{{\nu_1}{\nu_2}}$ are applied to each TT spectrum in the likelihood.
\npipe{} TT calibrations relative to $143 \times 143$ are given in Table \ref{tab:calibrations}. 

\subsubsection{Effective polarization efficiencies}
\begin{figure}
  \centering
  \includegraphics[width=0.8\columnwidth]{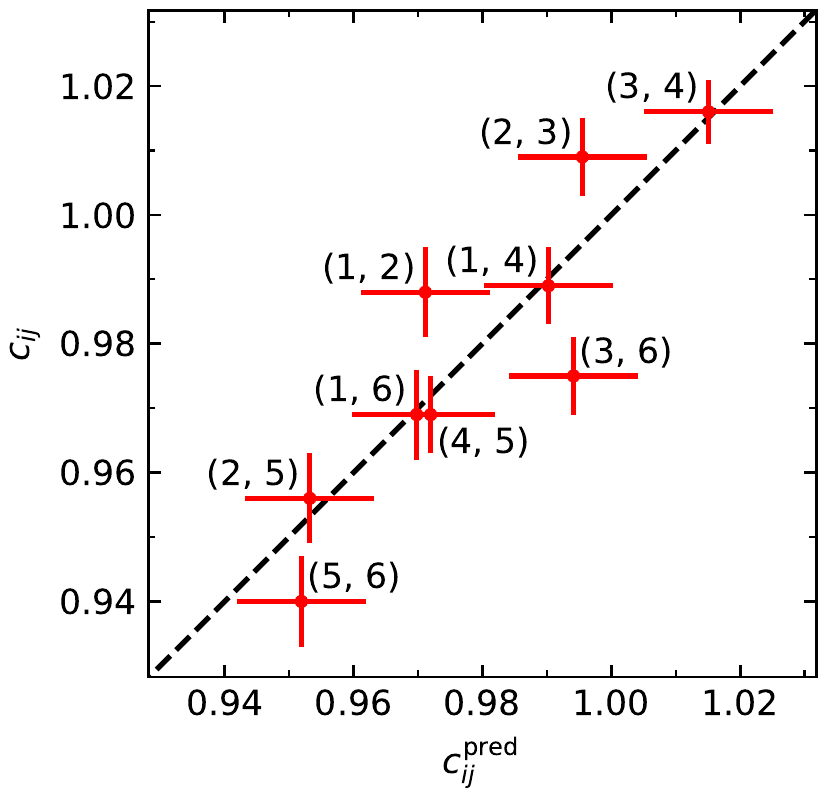}
  \caption{A comparison of \npipe{} 12.6 calibration factors $c_{ij}$ measured from the EE spectra and those predicted from triplets of TE spectra. The labels refer to the EE index given in Table \ref{tab:calibrations}.}
  \label{fig:map-spectrum-npipe}
\end{figure}

Calibration factors are also computed for the polarization spectra and interpreted as effective polarization efficiencies. In reality these may be due to a combination of real polarization efficiencies, polarization angle errors, and beam effects as in temperature. Similarly to temperature we calculate the calibration factor $c_k$ for a given TE or EE cross-spectrum $k$ by minimizing Eq.\ \ref{eq:calminpol}:
\begin{equation}
  \chi^2 = \sum_{\ell_1\ell_2}\left(C^k_{\ell_1} - c_kC_{\ell_1}^\mathrm{theory}\right)\left(M^k\right)^{-1}_{\,\,\ell_1\ell_2}\left(C^k_{\ell_2} - c_kC_{\ell_2}^\mathrm{theory}\right).
\label{eq:calminpol}
\end{equation}
Here $C_\ell^\mathrm{theory}$ is a theoretical power spectrum\footnote{$C_\ell^\mathrm{theory}$ here is the $\lcdm$ best-fit to \camspec{} 12.1HM TT from \eg.} and $M^k$ is the covariance matrix for spectrum $k$. We minimize with respect to $C_\ell^\mathrm{theory}$ to avoid instability in the estimate from the high noise in polarization; we expect any bias introduced favoring the theoretical model used for calibration to appear as an overall multiplicative factor that can be absorbed in the $c_\mathrm{TE}$ and $c_\mathrm{EE}$ factors of the likelihood (see Sec.\ \ref{sec:likelihood}). We see no indications for significant pulls outside the prior range of these calibration parameters or strong correlations with them in any $\lcdm$ extensions we test. The calibrations $c_k$, given in Table \ref{tab:calibrations}, are divided out from each spectrum before coaddition. As an additional check on these calibration factors, we can use the TE calibrations to predict those for EE. We group together TE spectra sharing the same polarization map and average their calibration factors to calculate an effective polarization efficiency $\bar{\rho}$ for each map. These can be combined into a predicted calibration $c_{ij}^\mathrm{pred} = \bar{\rho_i}\bar{\rho_j}$ for each EE spectrum. We compare these predicted values to the measured EE calibrations in Fig.\ \ref{fig:map-spectrum-npipe}. We see a strong correlation, suggesting that our interpretation of these calibration factors as effective polarization efficiencies is a good one.

\subsubsection{Temperature-to-polarization leakage}
In PR3 as well as \npipe{} great care was taken to eliminate temperature-to-polarization (TP) leakage due for example to bandpass mismatch or ADCNL. Another source of TP leakage is beam mismatch; since polarization maps are created using pairs of bolometers, any mismatch of the beams of the individual bolometers in a pair will create TP leakage in the map. We calculate a correction for this effect using the off-diagonal terms of the beam matrix $W_\ell$ computed using \quickpol{}, as shown in Eq.\ \ref{eq:leakage-matrix}:
\begin{subequations}
  \begin{align}    
  \bar{\bar{C}}_\ell^{TE} &= W_\ell^{TETE}\tilde{C}_\ell^{TE} + W_\ell^{TETT}\tilde{C}_\ell^{TT}\\
  \bar{\bar{C}}_\ell^{EE}& = W_\ell^{EEEE}\tilde{C}_\ell^{EE} + W_\ell^{EETT}\tilde{C}_\ell^{TT}.
    \end{align}
  \label{eq:leakage-matrix}
\end{subequations}
Here $\bar{\bar{C}}_\ell$ is the beam-convolved spectrum while $\tilde{C}_\ell$ is beam-corrected; in practice we use theoretical spectra $C_\ell^{TT}$ rather than $\tilde{C}_\ell^{TT}$ to estimate the leakage corrections. The \quickpol{} correction of Eq.\ \ref{eq:leakage-matrix} is applied for TE but found to be negligible for EE.

To test the effectiveness of the leakage correction, we follow the method laid out in \eg{} and organize the uncorrected TE spectra into `triplets'. Each triplet consists of three spectra all sharing the same polarization map but using different frequency maps in temperature. We expect TP leakage in the polarization map to correlate identically to the dominant CMB component of each temperature map, hence we look for correlated residuals in these triplets as signatures of TP leakage and compare to the correction factor of Eq.\ \ref{eq:leakage-matrix}. The TE spectra are mask and beam deconvolved, corrected for the polarization efficiencies of Table \ref{tab:calibrations}, and dust cleaned with 353\,GHz. Residuals are calculated with respect to the mean of all TE spectra rather than to a theoretical model to reduce the impact of cosmic variance.  
\begin{figure}
  \centering
  \includegraphics[width=\columnwidth]{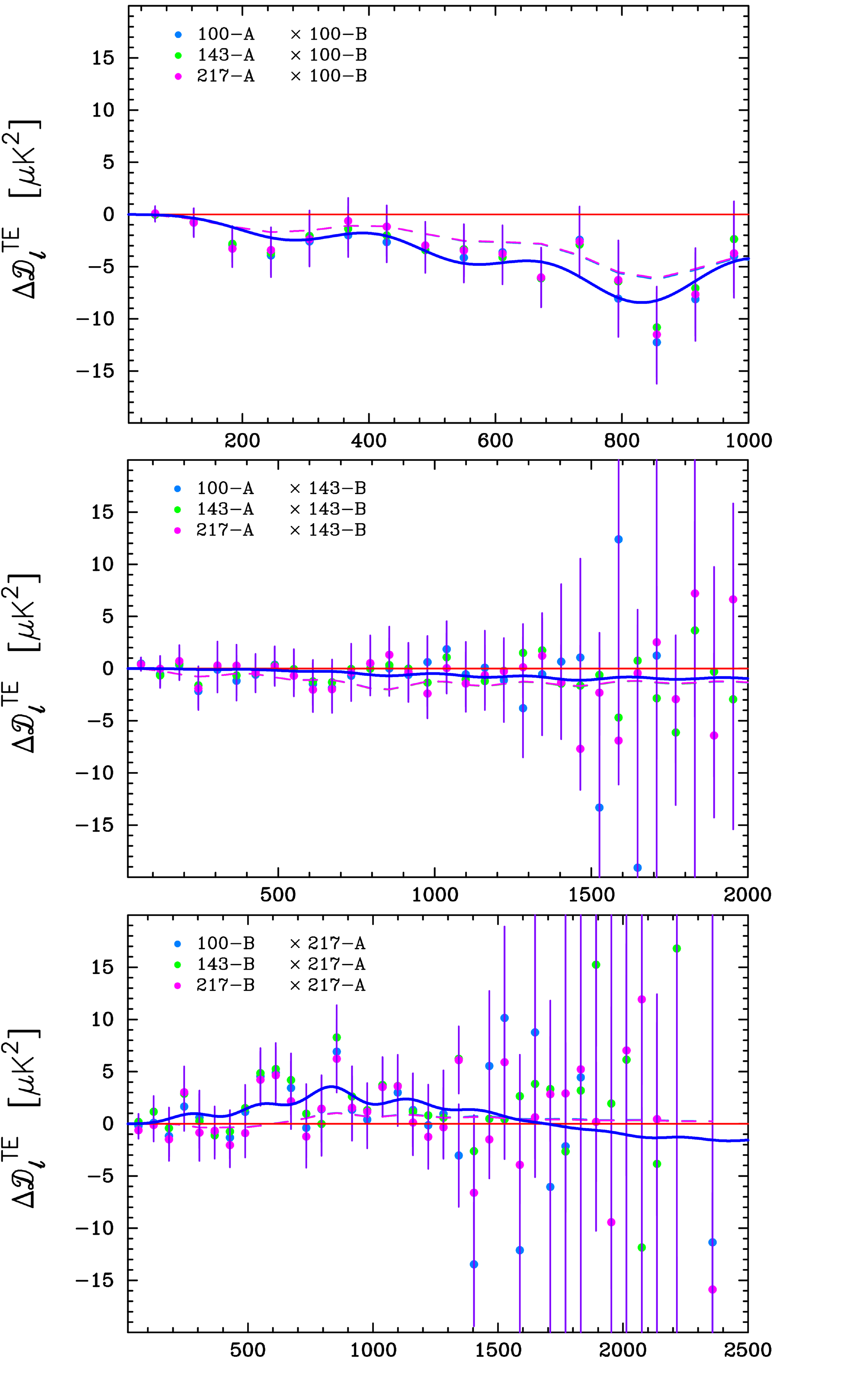}
  \caption{Residuals of \npipe{} TE spectra with respect to the mean of all TE spectra. The spectra are mask and beam deconvolved, 353\,GHz cleaned, and corrected for effective polarization efficiences. We see a clear correlation between spectra with the same polarization maps, suggesting the residuals are due to TP leakage. Dashed lines show the beam-induced leakage model (from Eq.\ \ref{eq:leakage-matrix}) corrected for in the likelihood. Solid lines are a fit to the residuals.}
  \label{fig:triplets-npipe}
\end{figure}

This test is shown for selected \npipe{} spectra in Fig.\ \ref{fig:triplets-npipe}; the solid lines are a fit to the data, while the dotted lines show the leakage correction derived from \quickpol{}. For 100\,GHz, where leakage is most significant, we see excellent agreement between the three spectra using different temperature maps, and between the data and the \quickpol{} prediction. The same is largely true of the other spectra as well, although there the leakage is quite small. Although the agreement is not perfect, complete neglect of this leakage is found in sec.\ 6.4 of \eg{} to cause up to $1\sigma$ shifts in cosmological parameters; therefore we expect any error from an imperfect leakage model to be significantly less than $1\sigma$. We also note that both the measured and predicted TP leakage is quite different between \npipe{} detector sets and the PR3 half-missions presented in \eg{}, being for example of a similar magnitude but opposite sign in 100\,GHz. Since the leakage correction is calculated from \quickpol{} in the same way in each case, we attribute this difference to the different split of the data in the two analyses.

\subsection{Noise Estimation and Covariances}
\label{sec:noise}
We use analytic covariance matrices as given in the appendix A.2 of \eg{}.
To model the noise in \planck{} we estimate a noise power spectrum for each map and fit it with an analytic function of the form
\begin{equation}
  N_\ell = A\left(\frac{100}{\ell}\right)^\alpha + B\frac{\left(\ell/1000\right)^\beta}{\left(1+\left(\ell/\ell_c\right)^\gamma\right)^\delta}.
  \label{eq:noisefn}
\end{equation}
Here $A$, $\alpha$, $B$, $\beta$, $\ell_c$, $\gamma$, and $\delta$ are free parameters. Weight functions $\psi_\ell = N_\ell / N_\ell^{\mathrm{white}}$ are computed for each $T$, $Q$, $U$ component, and the pixel noise estimates $(\sigma^X)^2_i$ in the original formulae for the covariances are replaced by $\sqrt{\psi^X_\ell \psi^X_{\ell^\prime}}(\sigma^X)^2_i$. 

Accurate modeling of the noise is especially important for the noise-dominated polarization spectra and has been a persistent challenge. Traditionally we have estimated noise spectra using differences of data splits. The \plik{} likelihood in \plkiii{} uses half-ring difference (HRD) maps for this purpose -- half-ring maps are made using the first or second halves of each pointing period and their difference gives an estimate of the noise. The PR3 half-ring maps are known to be correlated due to the glitch processing, so the HRD spectra are an underestimate of the true noise level. In \plkiii{} this effect was corrected for with an empirical bias correction; \eg{} chose instead to use odd-even ring differences (OED), calculated similarly to HRDs but using odd and even rings.  There is evidence that these maps overestimate the noise, particularly for $100 \times 100\GHz$ and at large scales, but in this paper we continue to use them for PR3. 
\begin{figure}
  \centering
  \includegraphics[width=\columnwidth]{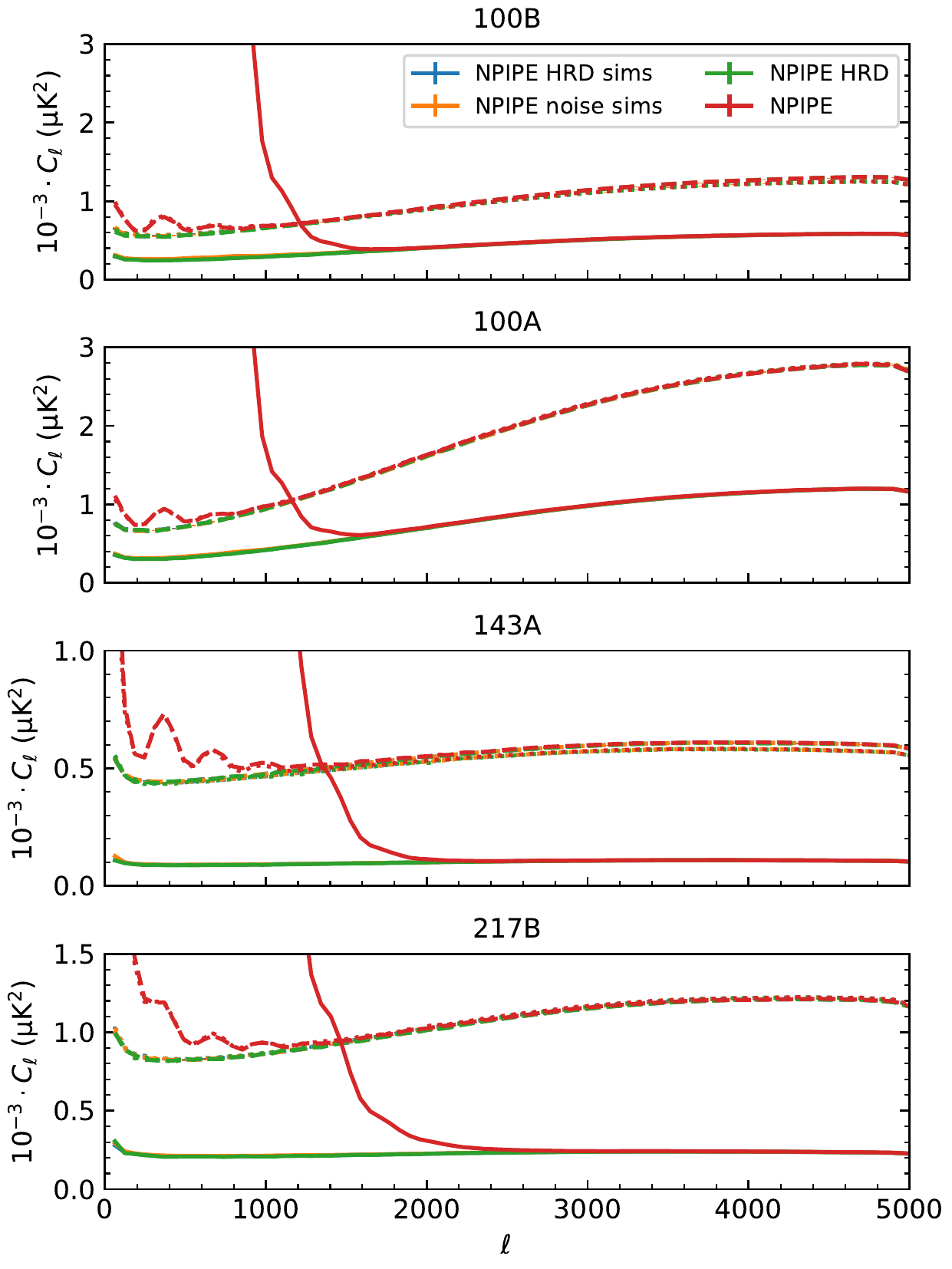}
  \caption{\npipe{} noise spectra before mask and beam deconvolution. Solid, dashed, and dotted lines show the auto-spectra of T, Q, and U maps respectively. For each frequency and split we show the auto-spectrum, half-ring difference (HRD) spectrum, and the simulated version of each. For the simulations we show the mean of 100 simulations. 143B and 217A are qualitatively similar to 143A/217B and are omitted.}
  \label{fig:npipenoisespectra}
\end{figure}
\begin{figure}
  \centering
  \includegraphics[width=\columnwidth]{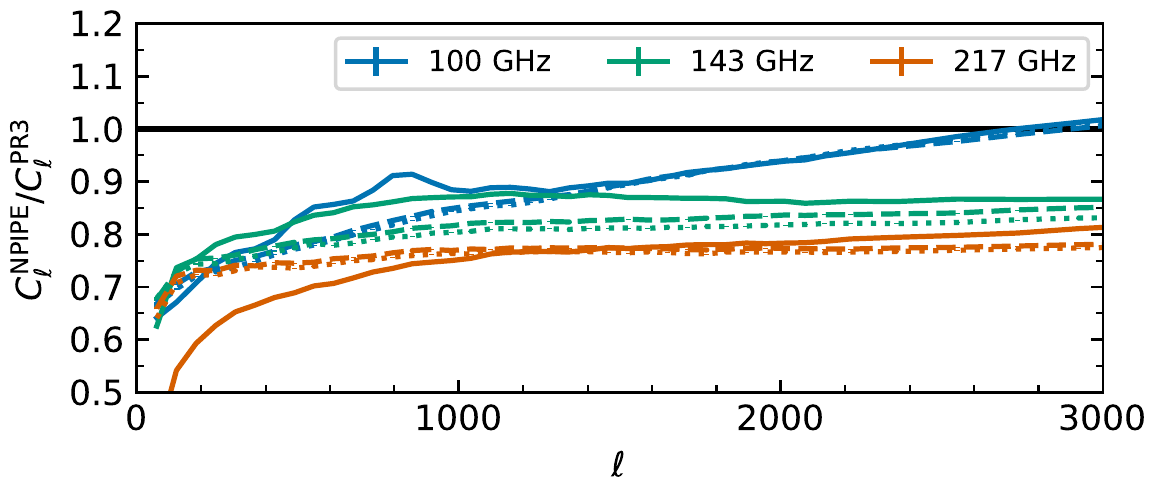}
  \caption{\npipe{} undeconvolved full-mission simulated noise spectra divided by their PR3 counterparts. Solid, dashed, and dotted lines show auto-spectra of T, Q, and U maps respectively.}
  \label{fig:npipe-vs-pr3-noise}
\end{figure}

In Fig.\ \ref{fig:npipenoisespectra} we present noise power spectra for \npipe{}, comparing noise estimates derived from simulations to auto-spectra of the detector-set maps and to HRDs. 
From all panels of Fig.\ \ref{fig:npipenoisespectra} we see no evidence of underestimation of the noise by \npipe{} HRDs at a level that would affect our analysis. \npp{} attributes this reduction in half-ring correlations to the changes in the glitch processing. We observe excellent agreement between the \npipe{} simulations, half-ring differences and auto-spectra at high multipoles. Below $\ell$ of 2000 however the half-ring difference spectra can be below the simulations by a few percent in polarization and up to 5\% in temperature, especially at 100\GHz. \npp{} also introduces a small-scale `noise alignment' correction to the simulated noise spectra in order to match to $\mathrm{A}-\mathrm{B}$ difference spectra. This further boosts the simulated noise power by 1-2\% in both temperature and polarization. These differences are of similar order to the accuracy of the fitting function Eq. \ref{eq:noisefn}, which is able to match the spectra to the 1-2\% level in polarization and 5\% in temperature. The current version of the \npipe{} likelihood also neglects to weight the half-ring difference map with per-pixel variances $w_i = \sigma_1 \sigma_2 / (\sigma^2_1+\sigma^2_2)$ (see eq. 5.3 of \eg{}), instead using a constant weighting of 0.5. This leads to a 10\% overestimate of the QQ noise for 100B and 143B and smaller (1-2\%) errors in polarization for other spectra. The effect on parameters should be small, but will be corrected in future work.

In Fig.\ \ref{fig:npipe-vs-pr3-noise} we see that \npipe{} has somewhat lower noise levels than the PR3 maps, potential overestimation from odd-even maps aside. This is expected to be due primarily to the addition of repointing data and the use of short baselines in destriping, and results in a 10-30\% noise reduction at the scales we use, dependent on frequency and multipole.

\subsection{Correlated Noise}
\label{sec:corrnoise}
\begin{figure}
  \centering
  \includegraphics[width=\columnwidth]{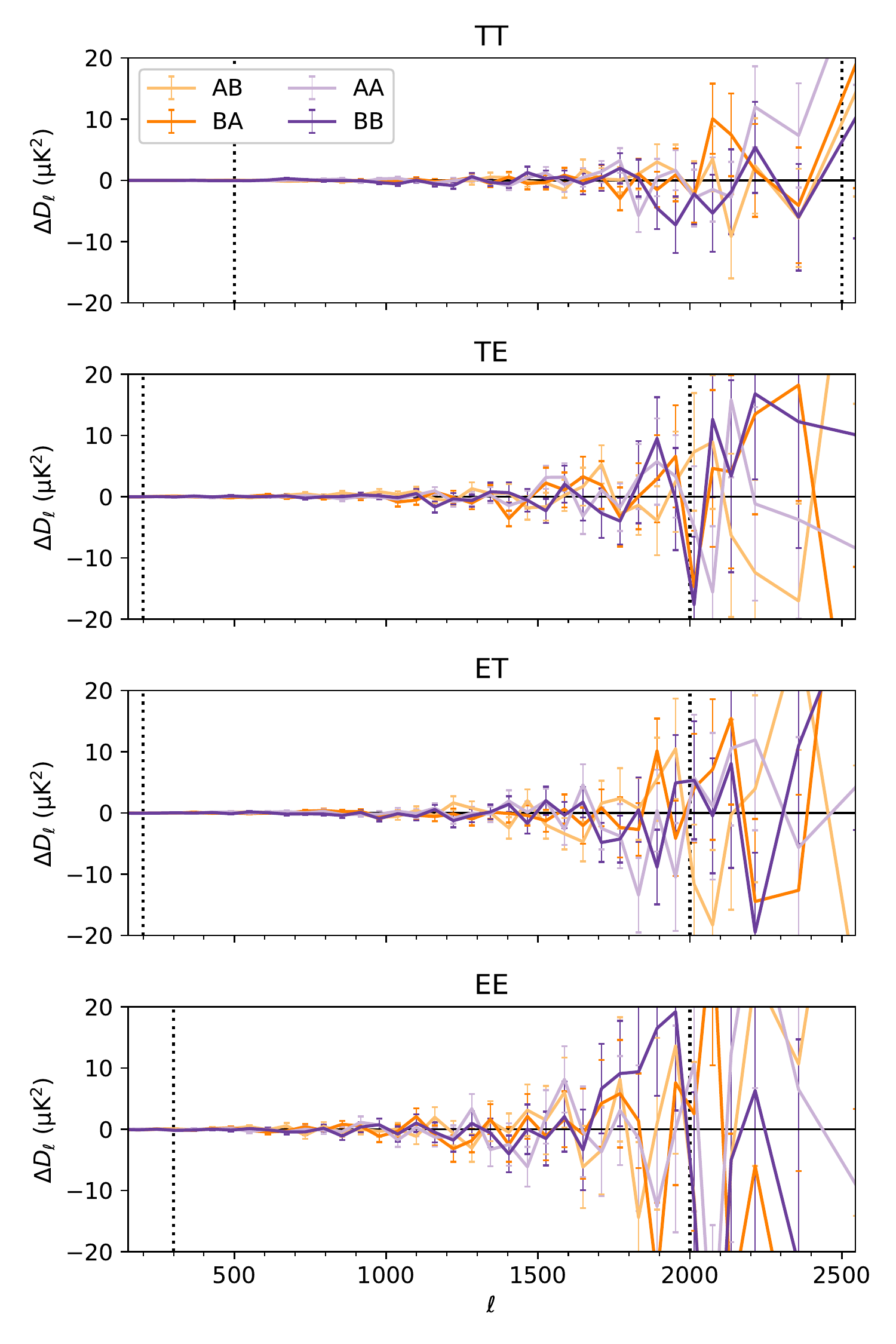}
  \caption{\npipe{} half-ring difference cross-spectra for $143\times 217\GHz$. Error bars show the standard deviation $\sigma / \sqrt{\Delta \ell}$ within each bandpower of bin-width $\Delta \ell$. Vertical lines indicate the $\ell$-range used in the likelihood.}
  \label{fig:npipecorrnoise}
\end{figure}

We use cross-spectra of the \npipe{} half-ring difference maps to check for correlated noise, with examples from $143\times 217$ shown in Fig.\ \ref{fig:npipecorrnoise} and others in Fig.\ \ref{fig:npipecorrnoisecross}. These spectra are consistent with zero, indicating there is negligible correlated noise; notably this is also true for $\mathrm{A}\times\mathrm{A}$ and $\mathrm{B}\times\mathrm{B}$ spectra. In particular we find that unlike for the PR3 detector sets the $143 \times 143$ and $217 \times 217$ TT spectra show no evidence of correlated noise. The most obvious deviations from zero are at high-$\ell$ in the $143\mathrm{B} \times 217\mathrm{B}$ EE spectrum, where the noise is nevertheless very high. Therefore we hope to add all AA and BB spectra in TE and EE to a future version of the likelihood as the addition of 6 spectra -- AA and BB for each of $100 \times 143$, $100 \times 217$, and $143 \times 217$ -- would significantly improve the constraining power of EE. TE would improve by a smaller amount as each new spectrum is equivalent to one already in the likelihood up to A-B exchange of the already signal-dominated temperature map.

\subsection{New likelihoods and nomenclature}
\label{sec:12_6}
Before discussing the spectra and parameter constraints in more detail we pause to introduce the nomenclature surrounding our likelihoods. As discussed previously our new likelihoods build on \camspec{}12.5HMcl, presented in \eg{}.  The covariance matrices of 12.5HMcl included contributions from foregrounds despite being a foreground-cleaned likelihood; this led to some overestimation of the error bars at high multipoles. This had negligible effect on cosmology but affected $\chi^2$ values in TT and has been corrected in the new 12.6HMcl, which in this paper we will call PR3\_12.6.  Using \npipe{} A and B maps in place of PR3 half-mission maps and half-ring instead of odd-even noise estimates we have an analogous \npipe{} likelihood, PR4\_12.6. 

\section{Spectra and Consistency Checks}
\label{sec:powspec}

\begin{figure}
  \centering
  \includegraphics[width=\columnwidth]{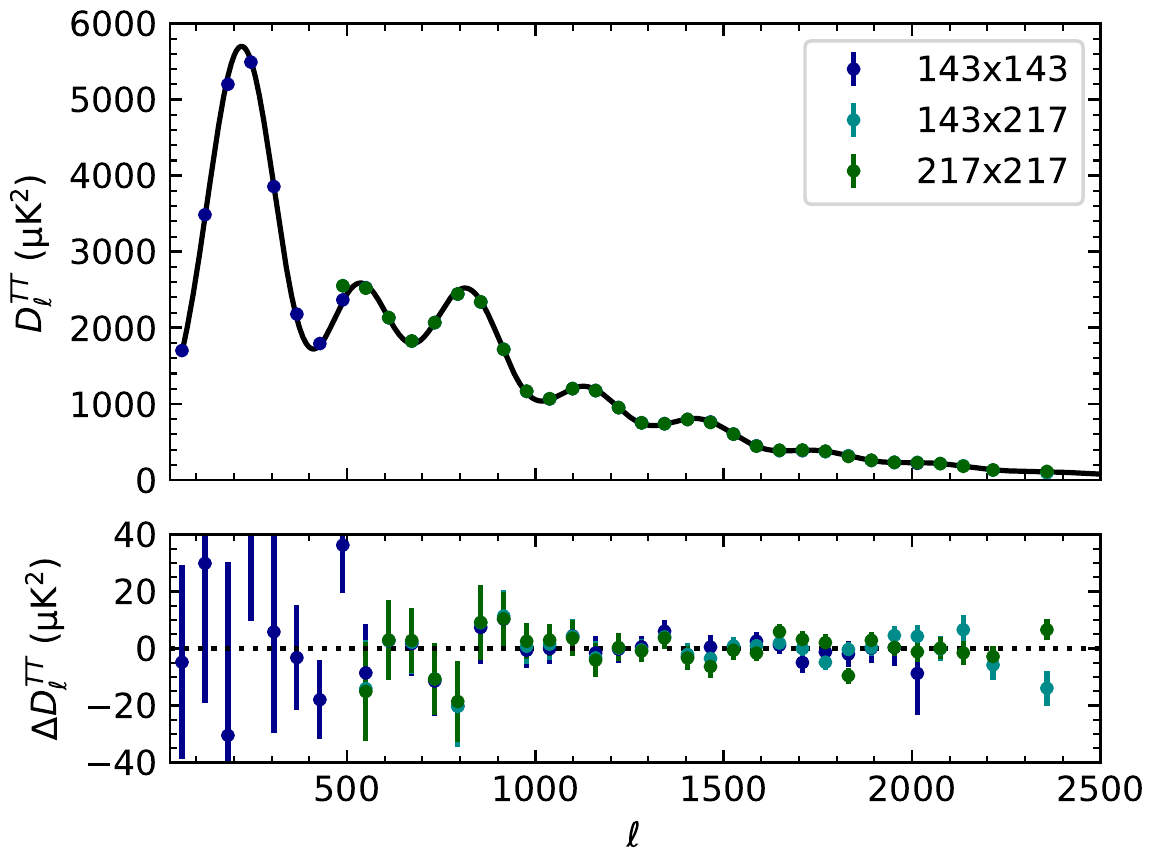}
  \caption{\npipe{} 12.6 TT power spectra with best-fit foregrounds removed. Residuals are shown with respect to the TTTEEE best-fit theory.}
  \label{fig:tt_powspec}
\end{figure}
\begin{figure}
  \centering
  \includegraphics[width=\columnwidth]{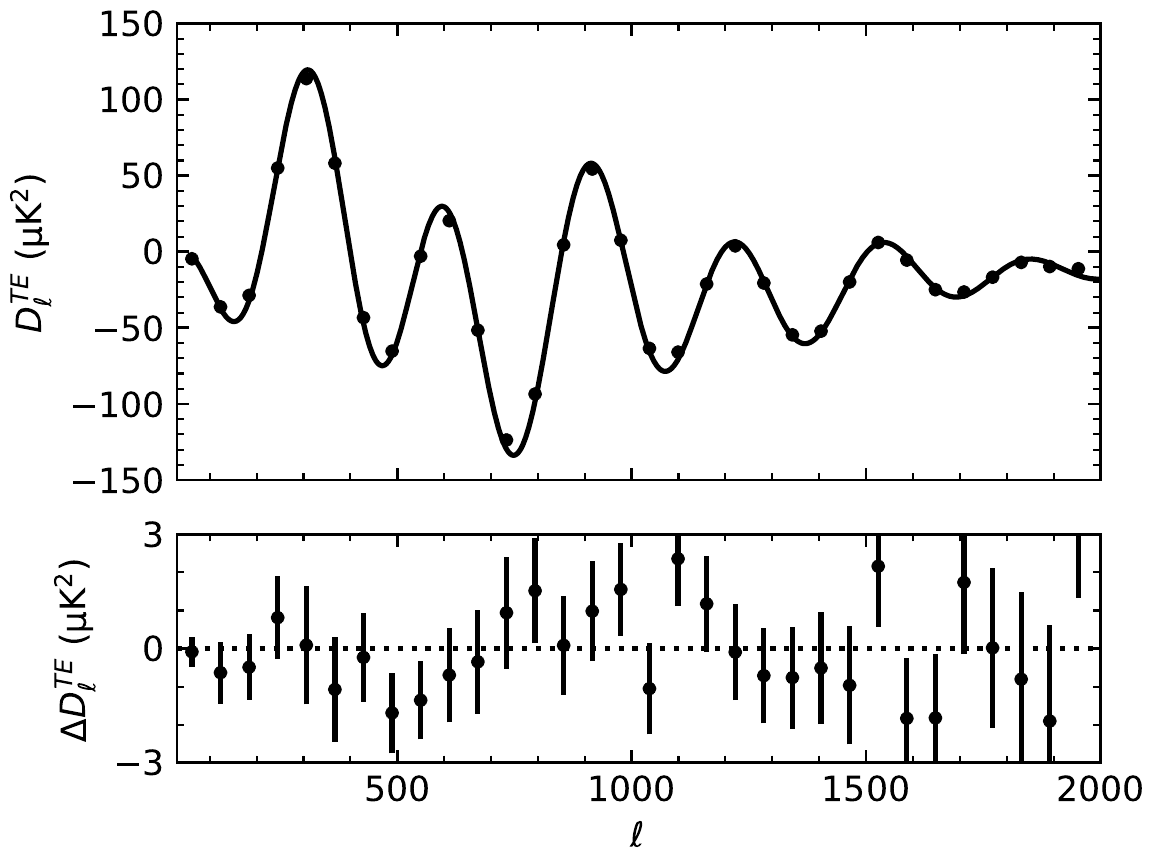}
  \caption{\npipe{} 12.6 TE power spectrum and residual with respect to the TTTEEE best-fit theory.}
  \label{fig:te_powspec}
\end{figure}
\begin{figure}
  \centering
  \includegraphics[width=\columnwidth]{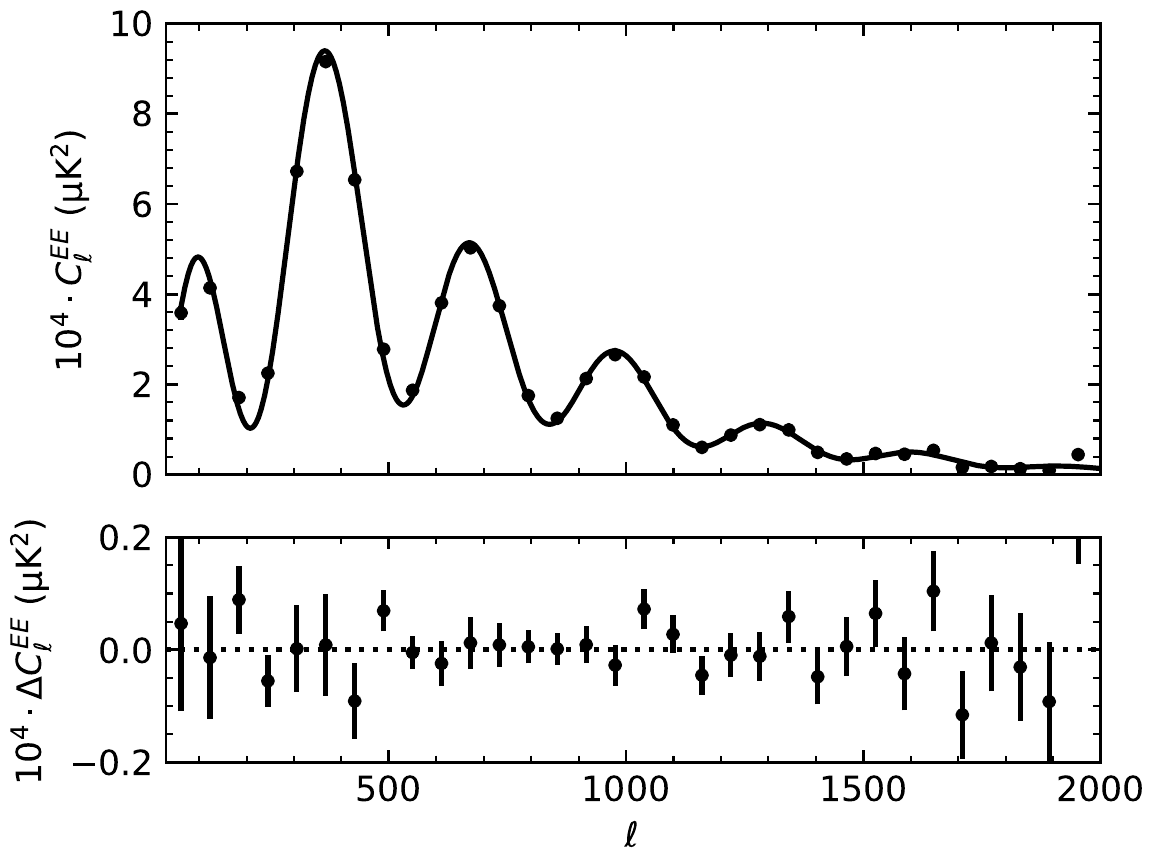}
  \caption{\npipe{} 12.6 EE power spectrum and residual with respect to the TTTEEE best-fit theory.}
  \label{fig:ee_powspec}
\end{figure}

\begin{figure}
  \centering
  \includegraphics[width=\columnwidth]{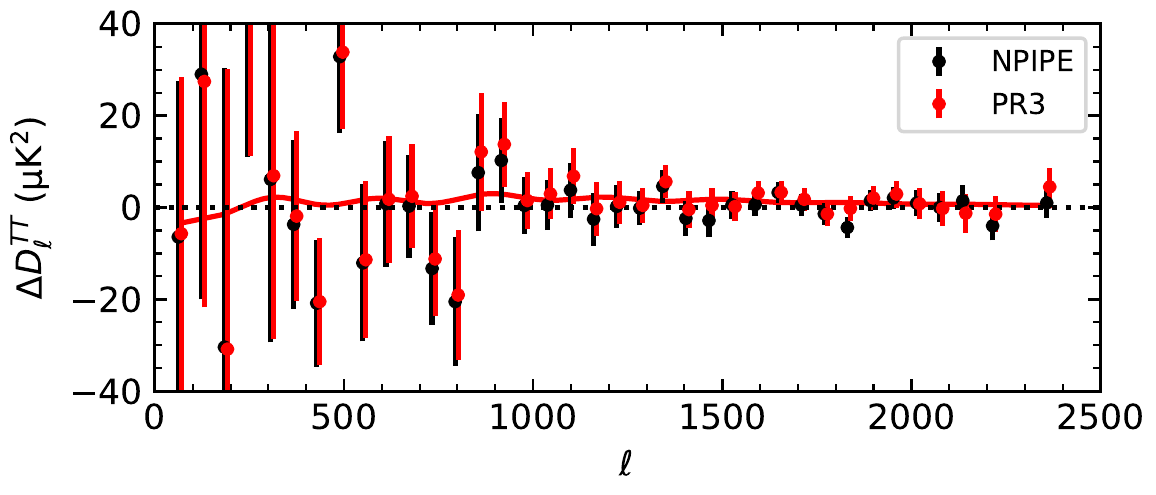}
  \caption{TT power spectrum residuals with respect to the PR4\_12.6 TT best-fit. Best-fit foreground residuals have been subtracted from the data points, while calibrations have been applied to the theories. The best-fit theory to PR3\_12.6 TT is shown as a solid red line. Points are slightly offset for visual clarity.}
  \label{fig:tt_powspec_compare}
\end{figure}

As in \camspec{} 12.5HMcl we use co-added spectra at $143\times143$, $143\times217$, and $217\times217$\GHz{} in TT, while co-adding all $\mathrm{A} \times \mathrm{B}$ cross-spectra of 100, 143, and 217\GHz{} in TE and EE. The multipole ranges used are given in Table \ref{tab:multipole}. Co-added \npipe{} 12.6 power spectra are shown in Figs.\ \ref{fig:tt_powspec} (TT), \ref{fig:te_powspec} (TE) and \ref{fig:ee_powspec} (EE), with residuals to the $\lcdm$ best-fit to TTTEEE (discussed in Sec.\ \ref{sec:likelihood}). Note that we plot binned data here for clarity, but the actual likelihood is unbinned. 

Visually the $\lcdm{}$ best-fit shown in Figs.\ \ref{fig:tt_powspec}-\ref{fig:ee_powspec} appears to match the spectra well; to more quantitatively check the agreement of the data and the model for PR4\_12.6 we calculate $\chi^2$ values with respect to the TTTEEE best-fit model, shown in Table \ref{tab:chi2}. For each of the individual TT spectra as well as co-added TT, TE and EE we find acceptable values for $\chi^2$. However, the $\chi^2$ values for total TT and TTTEEE are somewhat large, being over $4\sigma$ high. The source of these high $\chi^2$ is primarily at $500 \lesssim \ell \lesssim 1000$, and to a lesser extent $\ell < 500$. We see the same effect in PR3\_12.6, and it was also reported in \citet{planck2013-p08}. The effect is not unexpected as when all three temperature spectra are present and particularly when as now computed on the same galactic mask, the likelihood may respond to the `cosmic variance free' combination $C_\ell^{143\times143} + C_\ell^{217\times 217} - 2 C_\ell^{143\times 217}$ rendering the $\chi^2$ particularly sensitive to the accuracy of the noise modeling in these cases as the usually dominant cosmic variance is canceled. Comparing this combination to the expected null spectrum in the multipole range 500-2000 we find a $\chi^2$ high by $3.5\sigma$, supporting this interpretation. If we subtract off a version of this spectrum smoothed on a scale of $\Delta\ell = 61$ we still find that $\chi^2$ is high by $3.1\sigma$. This suggests we are seeing the effects of slightly high scatter throughout the $\ell$-range rather than a smooth trend that could be fit by an improved foreground model or a different cosmology.

At low multipoles (30-500) where we use only $143\times143$ the $\chi^2$ is $2.7\sigma$ higher than expected, but as there is only one spectrum this cannot be explained with the above argument. One possible explanation is the effect of excess variance introduced by the point-source masks that is not captured in the analytical covariance matrices. This effect is explained in depth in appendix C of \citet{planck2014-a13}. A deviation of up to 10\% in the variance is found with the masks used there; from that we estimate that this effect could account for up to about $1\sigma$ of the high variance we see at $\ell < 500$. For this region the improvement in $\chi^2$ is again marginal when a smoothed spectrum is subtracted, suggesting that relatively smooth cosmologies should not respond to this feature of the likelihood.

\begin{table}
\centering
\begin{tabular}{lccccc}
\hline
PR4\_12.6 & $\ell$ range & $N_D$ & $\hat \chi^2$ & $(\hat \chi^2 - 1) / \sqrt{2/N_D}$ \\
\hline
TT 143x143 & 30 -- 2000  & 1971  & 1.021 &  0.67 \\
TT 143x217 & 500 -- 2500 & 2001  & 0.985 & -0.47 \\
TT 217x217 & 500 -- 2500 & 2001  & 1.002 &  0.05 \\
TT Coadded & 30 -- 2500  & 2471  & 1.025 &  0.87 \\
TT All     & 30 -- 2500  & 5973  & 1.074 &  4.07 \\
TE         & 30 -- 2000  & 1971  & 1.055 &  1.73 \\
EE         & 30 -- 2000  & 1971  & 1.026 &  0.82 \\
TEEE       & 30 -- 2000  & 3942  & 1.046 &  2.02 \\
TTTEEE     & 30 -- 2500  & 9915  & 1.063 &  4.46 \\
\hline
\end{tabular}
\caption{$\chi^2$ of the different components of the PR4\_12.6 likelihood with respect to the TTTEEE best-fit model. $N_D$ is the size of the data vector. $\rchisq = \chi^2 / N_D$ is the reduced $\chi^2$. The last column gives the number of standard deviations of $\rchisq$ from unity.}
\label{tab:chi2}
\end{table}

To further investigate the internal consistency of the \npipe{} spectra, we can check how well the best $\lcdm$ fit to TT alone fits the TE and EE data. From Table \ref{tab:chi2-npipepol} we find reduced chi-squareds $\hat \chi^2_{TE} = 1.055$ and $\hat \chi^2_{EE} = 1.026$ using the \npipe{} TT best-fit solution. These values are $1.75\sigma$ and $0.83\sigma$ greater than 1 respectively, indicating that the TE and EE data are in acceptable agreement with the prediction of the TT model. We further compare these numbers to the best-fits to the TE and EE data to understand how close the TT model is to the actual minima of the polarization likelihoods. If we assume the TT model represents the truth, then due to over-fitting we expect the $\chi^2$ of the TE and EE minima to improve over the `true' (TT) model by a draw from a chi-squared distribution with $N_\mathrm{DOF}$ equal to the number of free parameters (see e.g.\ the appendix of \citet{grattonchallinor}). For TT-TE we find $\Delta \chi^2=2.4$, and for TT-EE $\Delta \chi^2=8.2$, roughly in line with the $\Delta \chi^2$ we expect for a 7-parameter model (6 $\lcdm$ parameters and one free calibration) and therefore giving us confidence that the \npipe{} polarization and temperature spectra are consistent.

\begin{figure}
  \centering
  \includegraphics[width=\columnwidth]{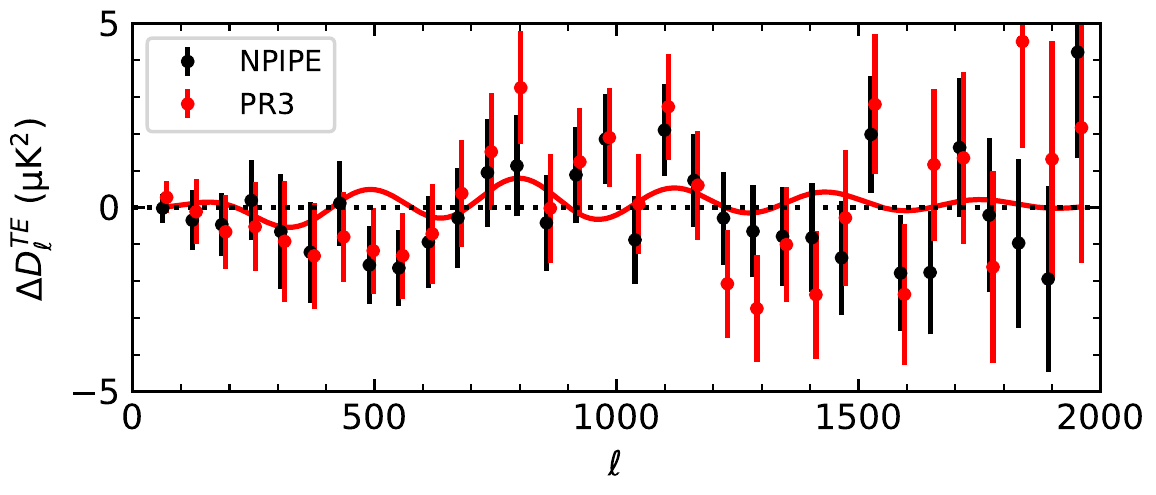}
  \caption{TE power spectrum residuals with respect to the PR4\_12.6 TE best-fit. The best-fit theory to PR3\_12.6 TE is shown as a solid red line.}
  \label{fig:te_powspec_compare}
\end{figure}

\begin{figure}
  \centering
  \includegraphics[width=\columnwidth]{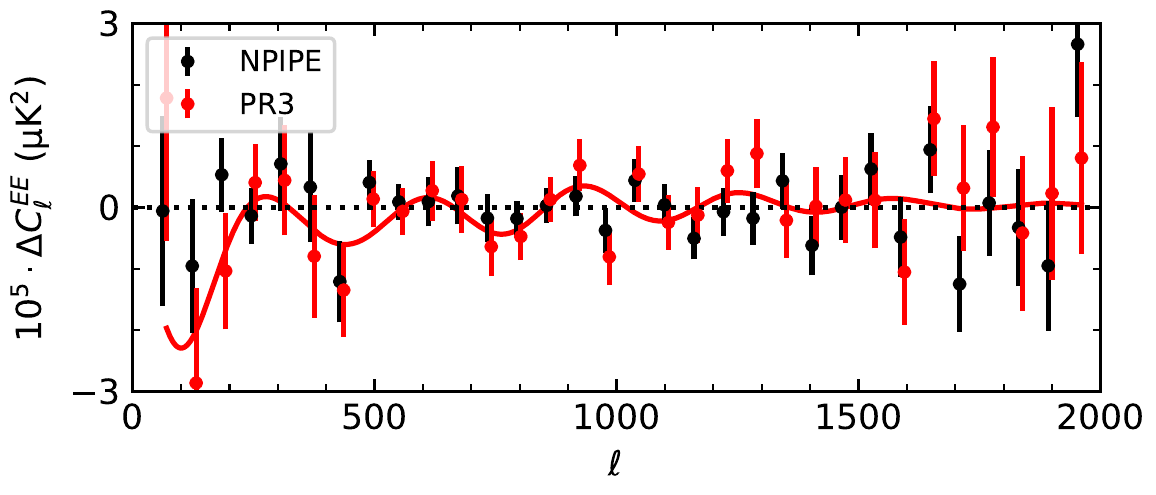}
  \caption{EE power spectrum residuals with respect to the PR4\_12.6 EE best-fit. The best-fit theory to PR3\_12.6 EE is shown as a solid red line.}
  \label{fig:ee_powspec_compare}
\end{figure}

Satisfied that our \npipe{} spectra are internally consistent, we compare them directly to the PR3\_12.6 spectra in Figs.\ \ref{fig:tt_powspec_compare}, \ref{fig:te_powspec_compare}, and \ref{fig:ee_powspec_compare}. Beginning with temperature in Fig.\ \ref{fig:tt_powspec_compare}, we find the co-added \npipe{} TT spectrum to be up to 0.3\% low relative to PR3 at $200 \lesssim \ell \lesssim 1600$\footnote{The small offset is due to a smooth difference of 2-5 $\muKsq$ in all three TT spectra in the given $\ell$-range. Similar such differences are discussed in connection with effective calibrations in sec.\ 6.2 of \eg{} for PR3 detector-set spectra.}.
This is reflected in very small cosmological parameter shifts in TT (see Table \ref{tab:allparams}).
In TE (Fig.\ \ref{fig:te_powspec_compare}) we see good agreement between the spectra apart from a few exceptional points, and find similar $\lcdm$ best-fits. The \npipe{} residuals are slightly smaller, especially at $\ell > 1000$, but with a higher $\chi^2$ due to the smaller error bars.
There is somewhat more difference between the EE spectra (Fig.\ \ref{fig:ee_powspec_compare}), with more large-scale power and a slight `wavy' residual that leads to a preference for lower $\thetastar$ in \npipe{} as will be seen in Fig.\ \ref{fig:npipe-comparepol} and Sec.\ \ref{sec:likelihood}.

To further quantify the level of agreement between the spectra we compare the PR4\_12.6 $\lcdm$ best-fit cosmology to the PR3\_12.6 data and vice-versa in Table \ref{tab:chi2-np-rd12}. In doing so we leave the nuisance parameters in the likelihood free, as they are not necessarily expected to be the same for each likelihood. We find $\Delta \chi^2 = 2.5$ between the PR3 and PR4 TTTEEE cosmologies for the PR4 data and $\Delta \chi^2 = 1.4$ for the PR3 data, indicating similar best-fit cosmologies compatible with both datasets; the agreement is similarly good for TT and TE individually. For EE we have $\Delta \chi^2 = 8.8$ with PR4 data and $\Delta \chi^2 = 6.6$ for PR3. As with the visual comparison of Fig.\ \ref{fig:ee_powspec_compare}, this again indicates some difference between PR3 and PR4 in EE.

\begin{table}
\centering
\begin{tabular}{lccc}
  \multicolumn{4}{c}{$\chi^2 (\Delta \sigma)$}\\
  \hline  
\textsl{Data}& TT Theory & TE Theory & EE Theory \\
\hline
\textsl{TE}   & 2080.7 (1.75) & 2078.3 (1.71) & 2119.1 (2.36) \\
\textsl{EE}   & 2023.2 (0.83) & 2027.7 (0.90) & 2015.0 (0.70) \\
\textsl{TEEE} & 4123.5 (2.04) & 4124.9 (2.06) & 4161.4 (2.47) \\
  \hline
\end{tabular}
\caption{Chi-squared values for \npipe{} polarization data. The rows give the chi-squared $\chi^2$ and the number of standard deviations from unity $\Delta \sigma = (\chi^2/N_D-1)/\sqrt{2/N_D}$ for the TE, EE, and combined TEEE datasets when compared to $\lcdm$ best-fit models to the TT, TE, and EE data (columns).}
\label{tab:chi2-npipepol}
\end{table}

\begin{table*}
  \centering
\begin{tabular}{lcccc}
\hline
$\chi^2 (\Delta \sigma)$ & $\mathrm{NP_d-NP_{th}}$ & $\mathrm{NP_d-PR3_{th}}$ & $\mathrm{PR3_d-NP_{th}}$ & $\mathrm{PR3_d-PR3_{th}}$\\
\hline
TT     & 6417.4  (4.066) & 6417.8  (4.070)  & 6335.0  (3.312) & 6334.9  (3.311)\\
TE     & 2078.3  (1.709) & 2078.8  (1.717)  & 1965.0  (-0.096) & 1964.4  (-0.105)\\
EE     & 2015.1  (0.702) & 2023.9  (0.843)  & 1811.9  (-2.534) & 1805.3  (-2.637)\\
TTTEEE & 10543.1 (4.460) & 10545.5 (4.477)  & 10116.1 (1.428) & 10114.7 (1.418)\\
\hline
\end{tabular}
\caption{$\chi^2$ comparisons between PR4\_12.6 (NP) and PR3\_12.6. Each row gives the spectrum used for both the data and the best-fit cosmology. The first column uses \npipe{} for both data and theory, the second compares \npipe{} data to the best-fit theory from PR3, and so forth. Note that only cosmological parameters are kept fixed in the theory, while nuisance parameters are left free.}
\label{tab:chi2-np-rd12}
\end{table*}


\section{Base $\Lambda$CDM Results}
\label{sec:likelihood}

In this section we describe our constraints on cosmological parameters in the base $\lcdm$ model using the \npipe{} versions of the \camspec{} likelihood.
We use the standard set of six cosmological parameters for $\lcdm$ given in Table \ref{tab:priors} with their priors. In addition to these cosmological parameters we have nuisance parameters $A^{\rm power}_{\nu}$ and $\gamma^{\rm power}_{\nu}$ for each TT spectrum, representing the amplitudes and spectral indices of power-law fits to remaining foreground residuals:
\begin{equation}
  D_\ell^{\rm power} = A^{\rm power}\left(\frac{\ell}{1500}\right)^{\gamma^{\rm power}}.
  \label{eq:fgpowerlaw}
\end{equation}
These nuisance parameters are necessary to capture excesses from CIB and point sources in temperature that are not removed by 545\GHz{} cleaning. Any small foreground residuals remaining after power-law subtraction are ignored in the spectra and covariances. There are also three calibration parameters by which the theory is divided in the likelihood: the overall calibration $A_\mathrm{Planck}^2$ and individual polarization calibrations $C_{\mathrm{TE}}$ and $C_{\mathrm{EE}}$.
Because all of our likelihoods are limited to $\ell \geq 30$ we supplement with low-$\ell$ ($2 \leq \ell \leq 29$) likelihoods from PR3; for all temperature likelihoods we include the \texttt{Commander} TT likelihood \citep{planck2016-l04} and for polarization we add the \texttt{SimAll} EE likelihood \citep{planck2016-l03}. 

\begin{table}
  \centering
  \begin{tabular}{ll}
    \hline
    & Prior limits\\
    \hline
    $\log(10^{10} \As)$ & 1.61-3.91\\
    $100 \thetamc$ & 0.5-10\\
    $\ns$ & 0.8-1.2 \\
    $\ombh$ & 0.005-1 \\
    $\omch$ & 0.001-0.99 \\
    $\taure$ & 0.01-0.8\\
    \hline\hline
    $A_{\mathrm{Planck}}$ & $\mu=1$; $\sigma=0.0025$\\
    $c_{\mathrm{TE}}$ & $\mu=1$; $\sigma=0.01$ \\
    $c_{\mathrm{EE}}$ & $\mu=1$; $\sigma=0.01$ \\    
    $A^{\rm power}_{143}$ & 0-50\\
    $A^{\rm power}_{217}$ & 0-50\\
    $A^{\rm power}_{\mathrm{143x217}}$ & 0-50\\
    $\gamma^{\rm power}_{143}$ & 0-5 \\
    $\gamma^{\rm power}_{217}$ & 0-5 \\
    $\gamma^{\rm power}_{\mathrm{143x217}}$ & 0-5 \\
    \hline
  \end{tabular}
  \caption{Prior ranges on sampled parameters in both PR3\_12.6 and PR4\_12.6. Gaussian priors with mean $\mu$ and standard deviation $\sigma$ are used for the calibrations, all other priors are flat. The TE and EE calibration parameters are not used in temperature-only likelihoods, and the foreground nuisance parameters $A^{\rm power}_{\nu}$ and $\gamma^{\rm power}_{\nu}$ are not used in polarization-only likelihoods.}
  \label{tab:priors}
\end{table}

\begin{figure*}
  \includegraphics[width=0.8\textwidth]{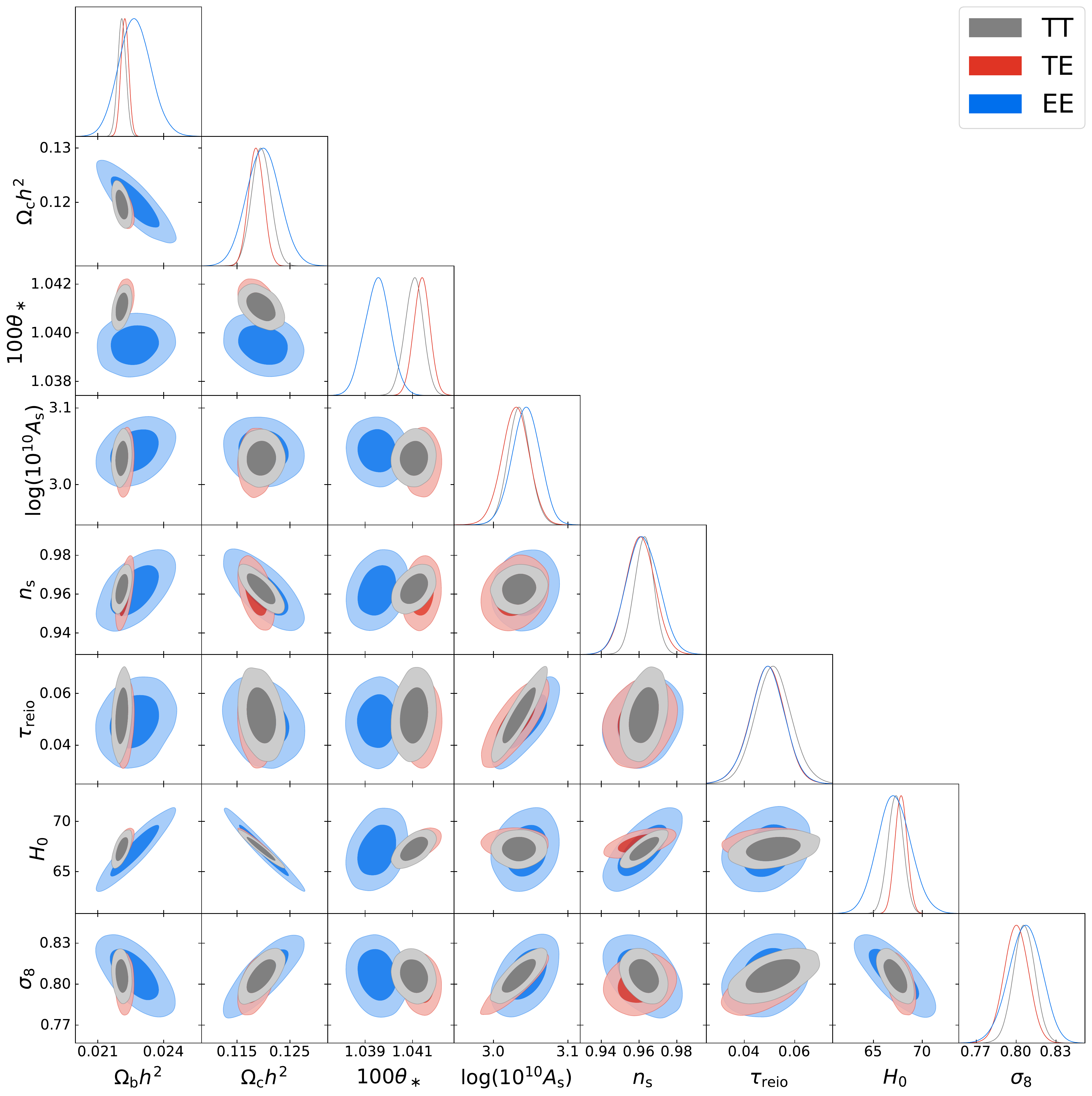}
  \caption{Comparison of 68\% and 95\% limits on cosmological parameters in $\lcdm$ from the \npipe{} TT, TE, and EE likelihoods.}
  \label{fig:npipe-comparepol}
\end{figure*}

In Fig.\ \ref{fig:npipe-comparepol} we compare parameter constraints from \npipe{} TT, TE and EE, again assessing agreement between the components. We find that the TT and TE constraints are highly consistent, and have similar constraining power; the EE constraints are broader but serve as a useful consistency check for \planck{}. Treating TT and EE as independent, we approximately quantify the consistency between their parameter constraints as follows. From our MCMC chains we compute a data vector of mean posterior values $P$ and covariances $C$ for our six sampled $\lcdm$ parameters. In this 6-dimensional space we calculate $\Delta_P = P_\mathrm{TT} - P_\mathrm{EE}$ and $C_\Delta = C_\mathrm{TT} + C_\mathrm{EE}$, finding $\chi^2 = \Delta_P^{\mathrm{T}}C_\Delta^{-1}\Delta_P = 7.31$, indicating agreement within 0.5$\sigma$. Indeed, EE is very consistent for all parameters except the acoustic scale parameter $\thetastar$, for which EE prefers a somewhat lower value (2.8$\sigma$ lower than TT).  This is reminiscent of the similar preference in PR3 EE for high $n_s$ relative to TT and TE which has disappeared in PR4\_12.6. The $\thetastar$ tension is also slightly worse for extended models: for $\lcdm + A_L$ we have a $3.3\sigma$ difference, and for $\lcdm + \omk$ $3.2\sigma$, due primarily to an increase in $\thetastar$ preferred by the TT spectrum in these models. Referring back to Fig.\ \ref{fig:npipe-comparepol} however (see also Fig.\ \ref{fig:hillipop-thetastar}), note that there is still substantial overlap of the $\thetastar$ contours; this is also true in the extended models. Given the consistency of \npipe{} TT and TE on $\thetastar$, which also agree very well with the values from PR3\_12.6, we do not regard this shift as evidence for lower $\thetastar$. Rather it is more likely due to parameter degeneracies coupling to residual systematics in EE (the spectrum most likely to be affected by systematics), or perhaps a statistical fluctuation.

\begin{figure*}
  \includegraphics[width=0.8\textwidth]{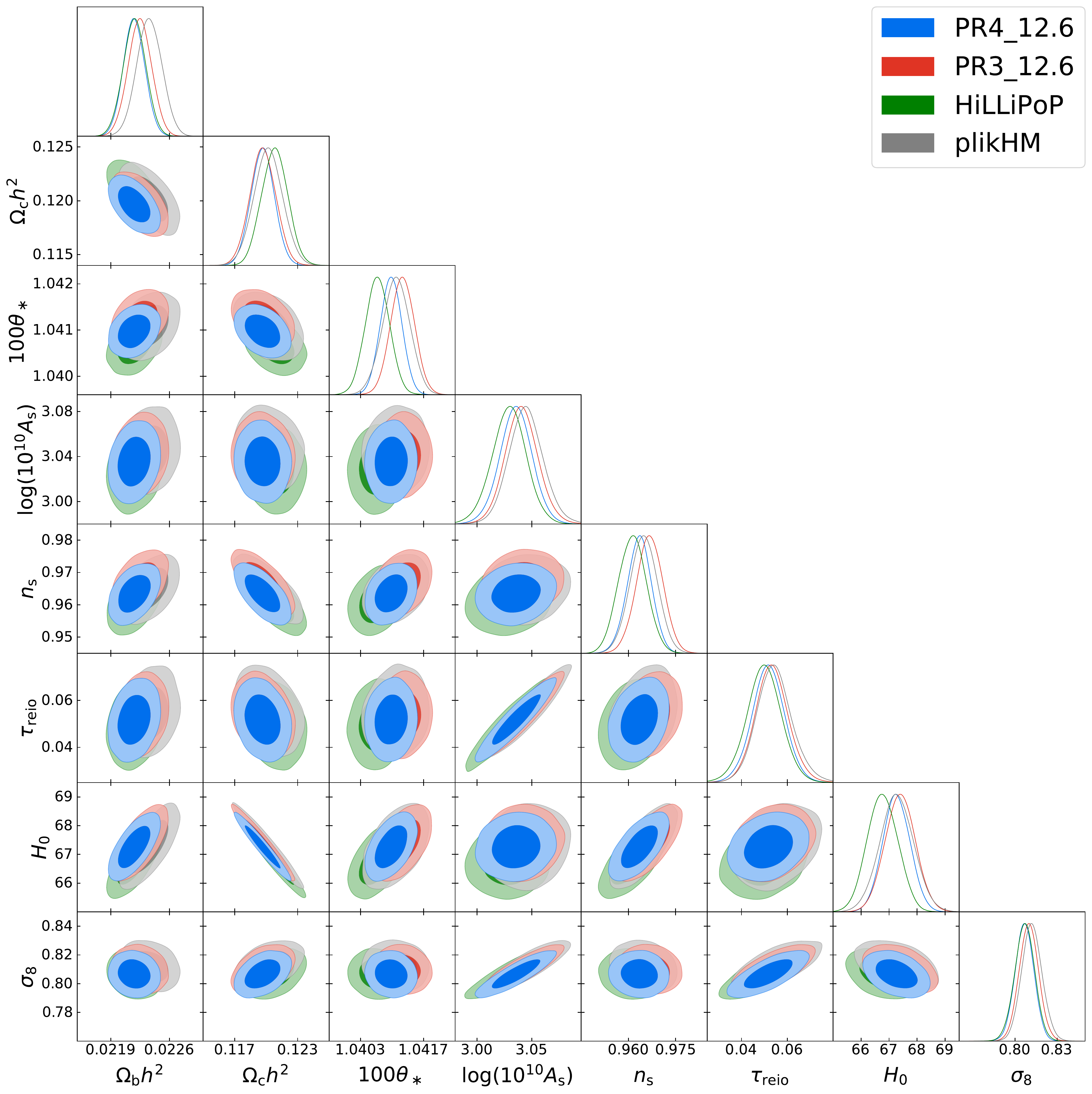}
  \caption{2D 68\% and 95\% parameter constraints on cosmological parameters in $\lcdm$ using the PR4\_12.6, PR3\_12.6, \texttt{HiLLiPoP} (PR4), and \plik{} (PR3) TTTEEE likelihoods. Note that the scales are zoomed in relative to Figure \ref{fig:npipe-comparepol}.}
  \label{fig:npipe-rd12-plik}
\end{figure*}

Moving on to the full TTTEEE likelihood, we present a comparison of $\lcdm$ parameter constraints from PR4\_12.6 and PR3\_12.6 in Table \ref{tab:npipeparams}. Two-dimensional marginalized constraints, alongside those from the 2018 \plik{} likelihood and \texttt{HiLLiPoP} (see Sec.\ \ref{sec:comparisons}), are given in Fig.\ \ref{fig:npipe-rd12-plik}. We find consistency at the sigma level among all four likelihoods, for all cosmological parameters. Given the small weight carried by EE in a TTTEEE likelihood, note that any issues with EE or differences in their handling will have little impact on the overall cosmological constraints. The \npipe{} likelihood contours are smaller than for PR3\_12.6, and significantly more constraining than \plik{}. The widths of the 1-dimensional 68\% constraints, as given in Table \ref{tab:npipeparams}, shrink between 7\% and 14\% (14-21\%) compared to PR3\_12.6 (\plik{}) TTTEEE for $\ombh$, $\omch$, $\thetastar$, $\ns$, and $H_0$. Such decreases are consistent with the lower noise levels discussed in Sec.\ \ref{sec:noise}.
\renewcommand{\tabcolsep}{2pt}
\begin{table}
\begin{tabular}{cccc}
\hline& PR4\_12.6& PR4\_12.6& PR3\_12.6\\
      & TT& TTTEEE& TTTEEE\\
\hline
$\Omega_\mathrm{b} h^2$                 &$0.02209\pm 0.00019$                    &$0.02218\pm 0.00013$                    &$0.02225\pm 0.00014$                    \\
$\Omega_\mathrm{c} h^2$                 &$0.1196\pm 0.0018$                      &$0.1197\pm 0.0011$                      &$0.1197\pm 0.0012$                      \\
$100\,\theta_\mathrm{MC}$               &$1.04083\pm 0.00039$                    &$1.04075\pm 0.00024$                    &$1.04102\pm 0.00027$                    \\
$\tau$                                  &$0.0515\pm 0.0073$                      &$0.0517\pm 0.0072$                      &$0.0533\pm 0.0074$                      \\
$\ln(10^{10} A_\mathrm{s})$             &$3.034\pm 0.015$                        &$3.035\pm 0.015$                        &$3.041\pm 0.015$                        \\
$n_\mathrm{s}$                          &$0.9626\pm 0.0052$                      &$0.9635\pm 0.0039$                      &$0.9667\pm 0.0042$                      \\
\hline
$H_0$                                   &$67.26\pm 0.79$                         &$67.26\pm 0.49$                         &$67.40\pm 0.54$                         \\
$100\,\thetastar$                       &$1.04107\pm 0.00038$                    &$1.04098\pm 0.00024$                    &$1.04123\pm 0.00027$                    \\
$\Omega_\Lambda$                        &$0.685\pm 0.011$                        &$0.6849\pm 0.0068$                      &$0.6859\pm 0.0076$                      \\
$\Omega_{\mathrm{m}}$                   &$0.315\pm 0.011$                        &$0.3150\pm 0.0068$                      &$0.3140\pm 0.0076$                      \\
$\Omega_{\mathrm{m}} h^2$               &$0.1423\pm 0.0017$                      &$0.1425\pm 0.0011$                      &$0.1426\pm 0.0012$                      \\
$\sigma_8$                              &$0.8060\pm 0.0082$                      &$0.8067\pm 0.0067$                      &$0.8098\pm 0.0070$                      \\
$\sigma_8 \Omega_{\mathrm{m}}^{0.5}$    &$0.452\pm 0.011$                        &$0.4528\pm 0.0072$                      &$0.4538\pm 0.0079$                      \\
$\sigma_8 \Omega_{\mathrm{m}}^{0.25}$   &$0.604\pm 0.010$                        &$0.6043\pm 0.0070$                      &$0.6062\pm 0.0075$                      \\
$z_\mathrm{re}$                         &$7.42\pm 0.76$                          &$7.43^{+0.75}_{-0.68}$                  &$7.58\pm 0.75$                          \\
$10^9 A_{\mathrm{s}}$                   &$2.079\pm 0.032$                        &$2.081\pm 0.031$                        &$2.092\pm 0.032$                        \\
$10^9 A_{\mathrm{s}} e^{-2\tau}$        &$1.875\pm 0.013$                        &$1.877\pm 0.011$                        &$1.880\pm 0.011$                        \\
$\mathrm{Age}/\mathrm{Gyr}$             &$13.826\pm 0.031$                       &$13.820\pm 0.020$                       &$13.806\pm 0.022$                       \\
$r_{\mathrm{drag}}$                     &$147.52\pm 0.43$                        &$147.40\pm 0.25$                        &$147.31\pm 0.28$                        \\
\hline
$A_\mathrm{Planck}$                        &$1.0004\pm 0.0024$                      &$1.0004\pm 0.0024$                      &$1.0005\pm 0.0024$                      \\
$c_\mathrm{TE}$                                &-                                       &$0.9975\pm 0.0036$                      &$1.0005\pm 0.0041$                      \\
$c_\mathrm{EE}$                                &-                                       &$0.9975\pm 0.0037$                      &$1.0013\pm 0.0043$                      \\
$A^{\rm power}_{143}$                   &$19.1^{+2.0}_{-2.8}$                    &$18.1^{+1.7}_{-2.4}$                    &$17.2^{+1.9}_{-2.4}$                    \\
$A^{\rm power}_{143\times217}$          &$9.8^{+1.9}_{-2.7}$                     &$8.8^{+1.6}_{-2.3}$                     &$8.0^{+1.7}_{-2.4}$                     \\
$A^{\rm power}_{217}$                   &$12.9^{+1.9}_{-2.7}$                    &$11.9^{+1.6}_{-2.3}$                    &$12.8^{+1.7}_{-2.4}$                    \\
$\gamma^{\rm power}_{143}$              &$0.97^{+0.23}_{-0.17}$                  &$1.01^{+0.20}_{-0.17}$                  &$1.12^{+0.28}_{-0.24}$                  \\
$\gamma^{\rm power}_{143\times217}$     &$1.54^{+0.53}_{-0.61}$                  &$1.72^{+0.54}_{-0.61}$                  &$1.53^{+0.51}_{-0.74}$                  \\
$\gamma^{\rm power}_{217}$              &$1.39\pm 0.42$                          &$1.52\pm 0.41$                          &$1.40\pm 0.39$                          \\
\hline
\end{tabular}
\caption{PR4\_12.6 parameter constraints in $\lcdm$, with a comparison to PR3\_12.6. We report mean values and 68\% confidence intervals. Further entries are given in Table \ref{tab:allparams}.}
\label{tab:npipeparams}
\end{table}
\renewcommand{\tabcolsep}{6pt}

\section{Extended Models}
\label{sec:extensions}

We turn now to constraints of single-parameter extensions to the $\lcdm$ model. In particular we have varied curvature $\omk$, the phenomenological $A_L$ parameter, the effective number of light species $\neff$, and the sum of the neutrino masses $\mnu$ one at a time. Marginalized constraints on each of these parameters are presented in Table \ref{tab:extended-params}, but we focus our discussion on $\omk$ and $A_L$.

\begin{table}
  \centering
  \begin{tabular}{|c|c|c|c|c|}
    \hline
    PR4\_12.6 & $A_L$                     & $\omk$ & $\neff$ & $\mnu$ \\
    \hline
    TTTEEE& $1.095 \pm 0.056$         & $-0.025^{+0.013}_{-0.010}$ & $3.00 \pm 0.21$ & $ < 0.161$ \\
    TT    & $1.198 \pm 0.084$         & $-0.042^{+0.022}_{-0.016}$ & $2.98^{+0.28}_{-0.35}$ & $ < 0.278$ \\
    TE    & $0.96 \pm 0.15$           & $-0.010^{+0.035}_{-0.015}$ & $3.11^{+0.38}_{-0.42}$ & $ < 0.400$ \\
    EE    & $0.995 \pm 0.15$          & $-0.012^{+0.034}_{-0.017}$ & $4.6 \pm 1.3         $ & $ < 2.37 $ \\
    \hline
    PR3\_12.6 & $A_L$                     & $\omk$ & $\neff$ & $\mnu$ \\
    \hline
    TTTEEE& $1.146 \pm 0.061$         & $-0.035^{+0.016}_{-0.012}$ & $2.94^{+0.20}_{-0.23}$ & $ < 0.143$ \\
    TT    & $1.215 \pm 0.089$         & $-0.047^{+0.024}_{-0.017}$ & $2.89^{+0.28}_{-0.32}$ & $ < 0.248$ \\
    TE    & $0.96 \pm 0.17$           & $-0.015^{+0.043}_{-0.015}$ & $2.96^{+0.42}_{-0.49}$ & $ < 0.504$ \\
    EE    & $1.15 \pm 0.20$           & $-0.053^{+0.063}_{-0.029}$ & $2.46^{+0.94}_{-1.7} $ & - \\
    \hline
  \end{tabular}
  \caption{Mean values and 68\% limits for beyond-$\lcdm$ parameters.}
  \label{tab:extended-params}
\end{table}

\begin{figure*}
  \centering
  \includegraphics[width=0.35\textwidth]{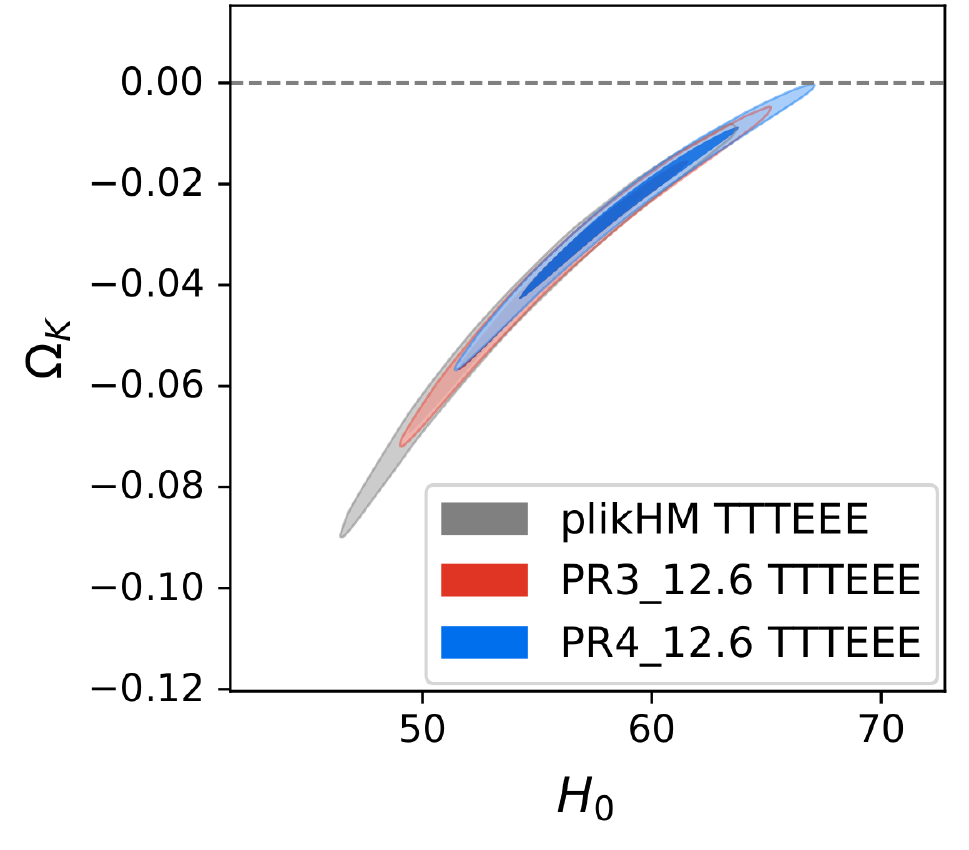}
  \includegraphics[width=0.35\textwidth]{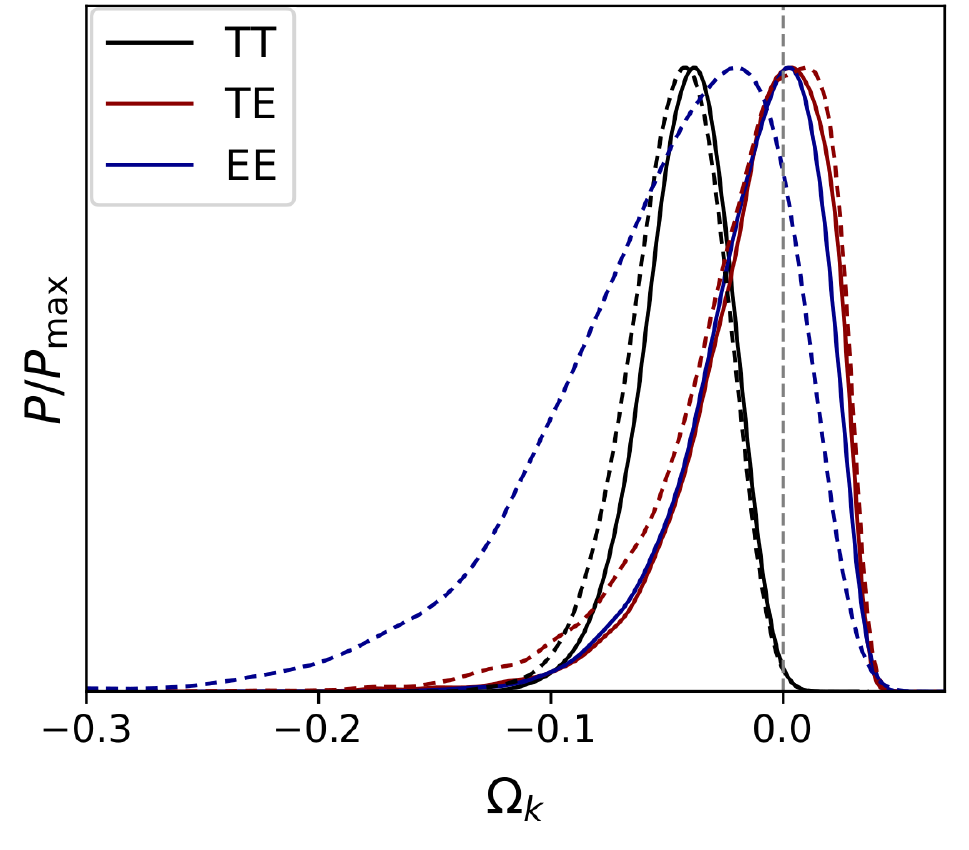}
  \caption{Left: 68\% and 95\% confidence limits in the $\omk-H_0$ plane from \plik{}, PR3\_12.6, and PR4\_12.6 TTTEEE likelihoods. Right: 1D marginalized densities on $\omk$ from TT, TE, and EE. Constraints from \npipe{} are shown with solid lines, and PR3\_12.6 with dashed lines.}
  \label{fig:omk}
\end{figure*}

Starting with $\omk$ we see a notable difference between \npipe{} and PR3 polarization. The TT results hardly change, but in TE and especially EE we find with \npipe{} a significant shrinking of the error bars, accompanied by a shift toward $\omk = 0$, as illustrated in Fig.\ \ref{fig:omk}. In particular, the long tails allowing negative $\omk$ are significantly reduced. Inspection of the 2D contours (not illustrated) do not suggest any obvious degeneracy between $\omk$ and $\thetastar$, so this shift in $\omk$ appears independent from the low value of $\thetastar$ seen above in \npipe{} EE. The maximum-likelihood point (best-fit) of PR4\_12.6 TTTEEE prefers $\omk < 0$ at $1.2\sigma$, similar to the $1.6 \sigma$ preference in PR3\_12.6 but now with a shift of the best-fit from $\omk = -0.026$ to $-0.016$  and a 20\% decrease in the size of the positive error bar. Similarly, the mean value of the posteriors prefers $\omk < 0$ at $1.9\sigma$ in PR4\_12.6 TTTEEE, down from $2.2\sigma$ in PR3 (see Table \ref{tab:extended-params}).

\begin{figure}
  \centering
  \includegraphics[width=0.7\columnwidth]{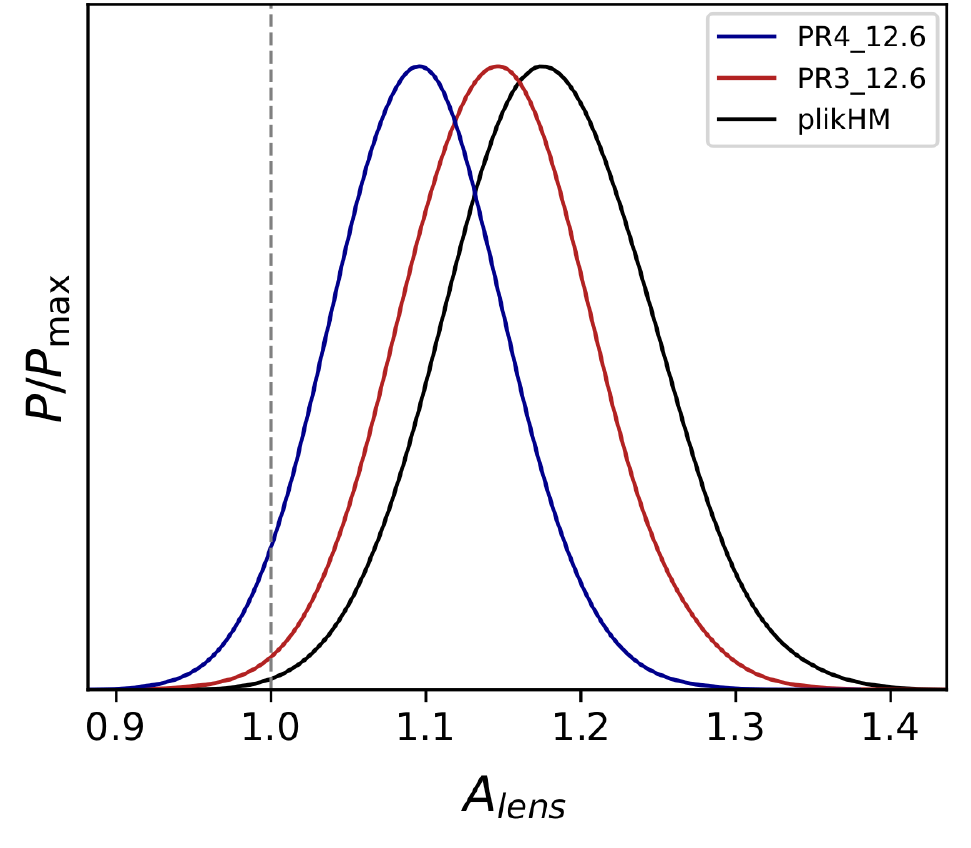}
  \caption{Marginalized constraints on $A_L$ from the PR4\_12.6, PR3\_12.6, and \texttt{Plik} TTTEEE likelihoods.}
  \label{fig:alens-1d}
\end{figure}

Turning to $A_L$, shown in Fig.\ \ref{fig:alens-1d}, we see a similar trend to that for $\omk$. PR4\_12.6 prefers a lower best-fit value of $A_L = 1.084$ compared to $A_L = 1.141$ from PR3\_12.6, and is now $1.5\sigma$ greater than $A_L = 1$ rather than $2.3\sigma$. The means (Table \ref{tab:extended-params}) find $A_L > 1$ by $1.7\sigma$ (PR4) and $2.4\sigma$ (PR3)\footnote{Considering the data entering PR3\_12.6 as a subset of PR4\_12.6 then one can apply the method of \citet{grattonchallinor} to estimate the expected standard deviation of the shifts to be $\sqrt{0.061^2 - 0.056^2} = 0.024$, putting the difference of the means at 2.1 sigma.}. The constraint from EE in particular has shifted to be centered at $A_L=1$, while TT and TE show smaller downward shifts with slightly smaller error bars relative to PR3\_12.6. Again, there does not appear to be a particular degeneracy between $A_L$ and $\thetastar$. To better test this we re-run cosmological parameter estimation with $\thetastar$ fixed to its PR3\_12.6 best-fit value, and find that the \npipe{} $A_L$ constraints are effectively unchanged. This gives us confidence that the preference for excess smoothing of the power spectra represented by $A_L$ really has decreased in the \npipe{} maps, independent of other parameter shifts.

\section{External Comparisons}
\label{sec:comparisons}
\begin{figure}
  \centering
  \includegraphics[width=0.7\columnwidth]{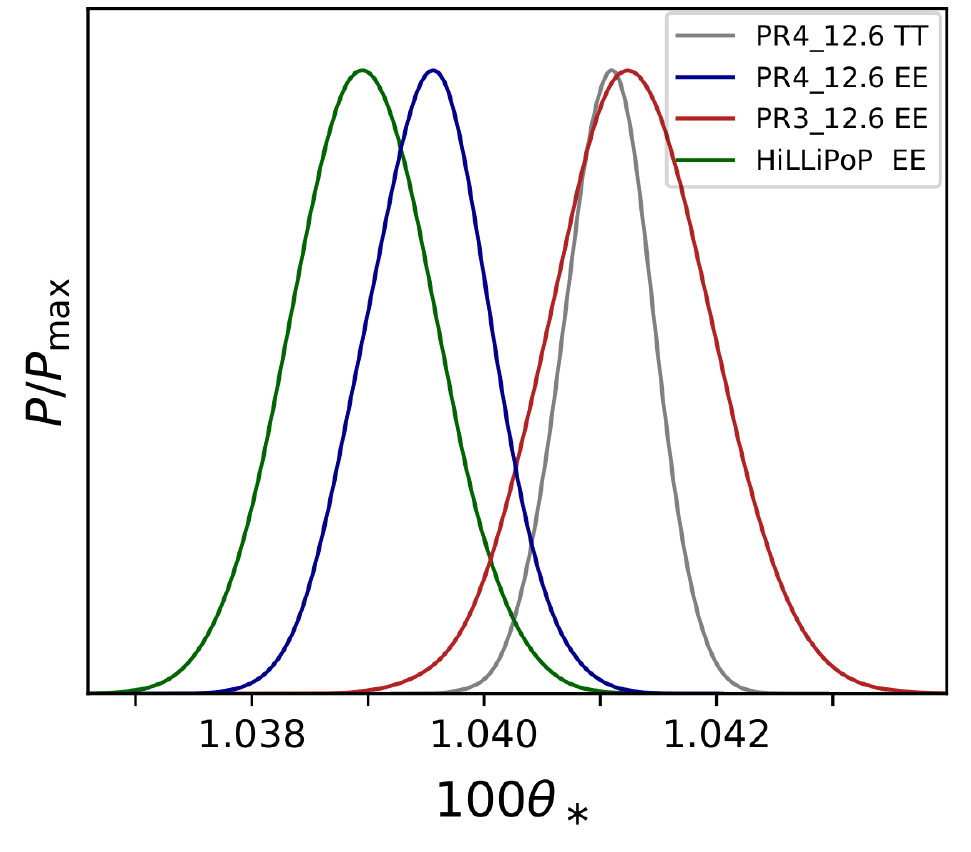}  
  \caption{Marginalized constraints on $\thetastar$ from PR4\_12.6 TT and the PR4\_12.6, PR3\_12.6, and \hillipop{} (PR4) EE likelihoods. Both \npipe{} EE likelihoods prefer lower values of $\thetastar$ relative to PR3 EE or TT.}
  \label{fig:hillipop-thetastar}
\end{figure}

\begin{figure*}
  \centering
  \includegraphics[width=\textwidth]{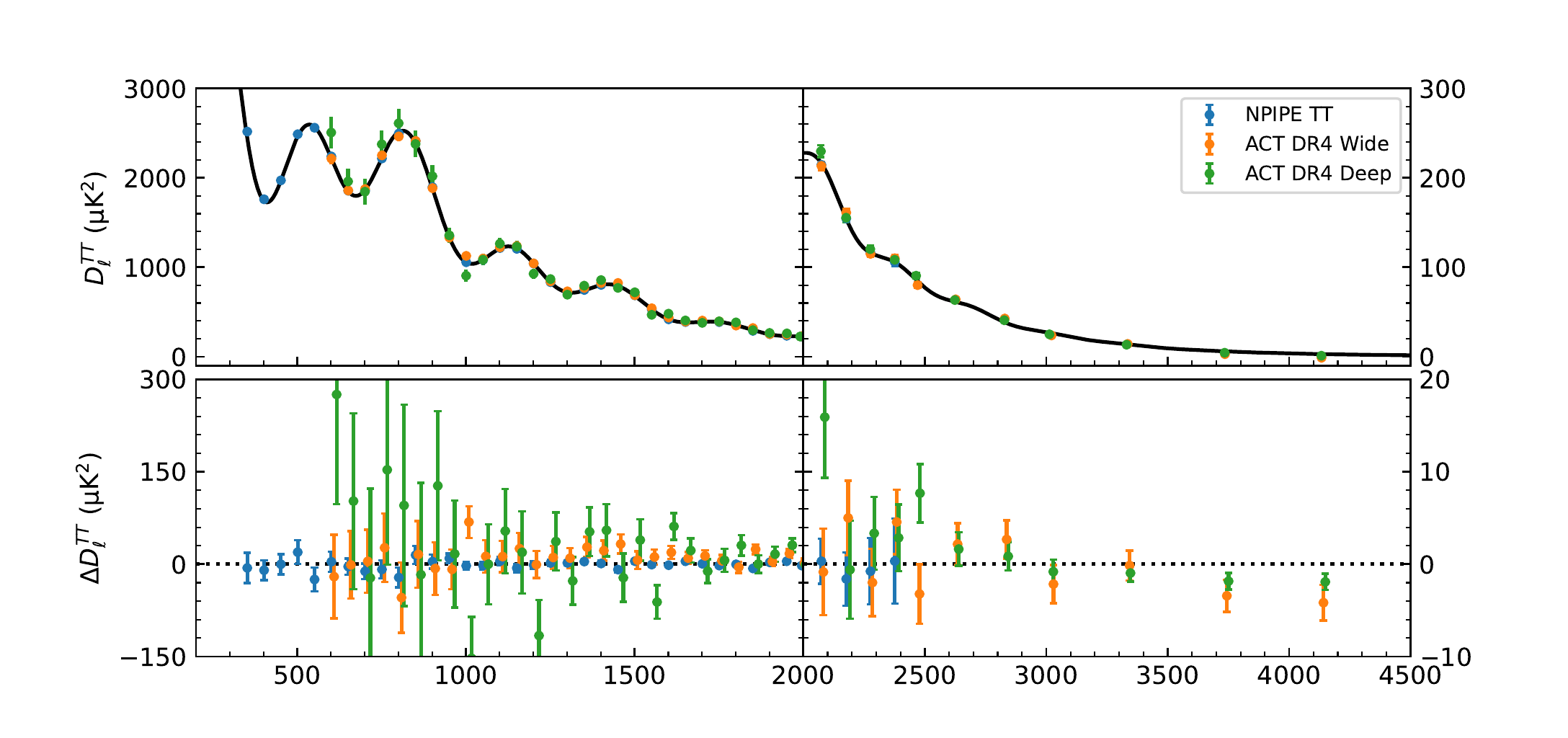}
  \caption{PR4\_12.6 and ACT DR4 TT power spectra. \npipe{} spectra are foreground-subtracted and binned using the ACT Wide window functions. The lower panels show residuals to the best-fit PR4\_12.6 TT power spectrum. At multipoles higher than 2000 we use an expanded scale to better show differences between the spectra.}
  \label{fig:actTT}
\end{figure*}
\begin{figure*}
  \centering
  \includegraphics[width=\textwidth]{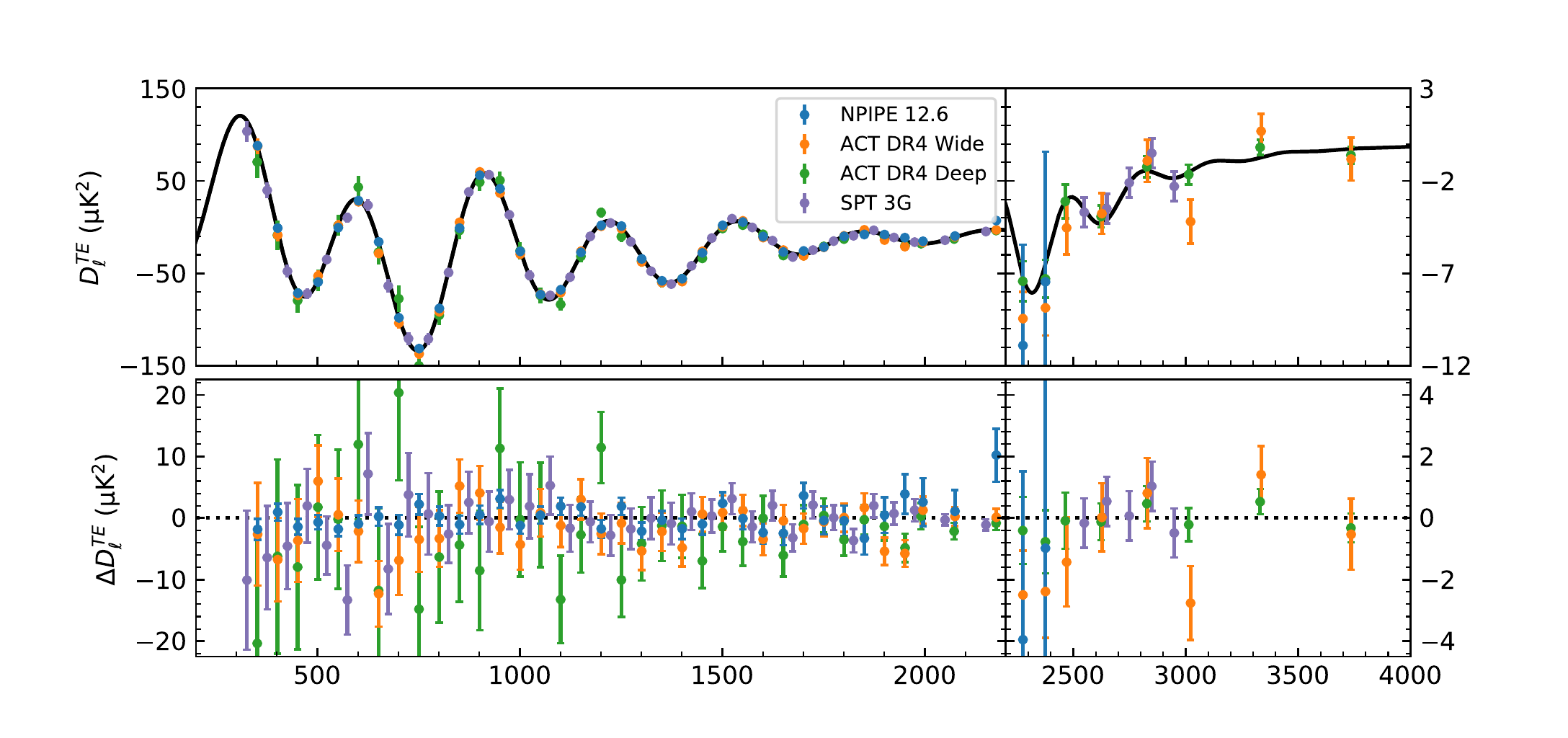}
  \caption{PR4\_12.6, ACT DR4, and SPT-3G TE power spectra. The lower panels show residuals to the best-fit PR4\_12.6 TE power spectrum, binned using the appropriate window function. \npipe{} is binned as ACT Wide. As in Fig. \ref{fig:actTT} the scale is expanded at high $\ell$.}
  \label{fig:actsptTE}
\end{figure*}
\begin{figure*}
  \centering
  \includegraphics[width=\textwidth]{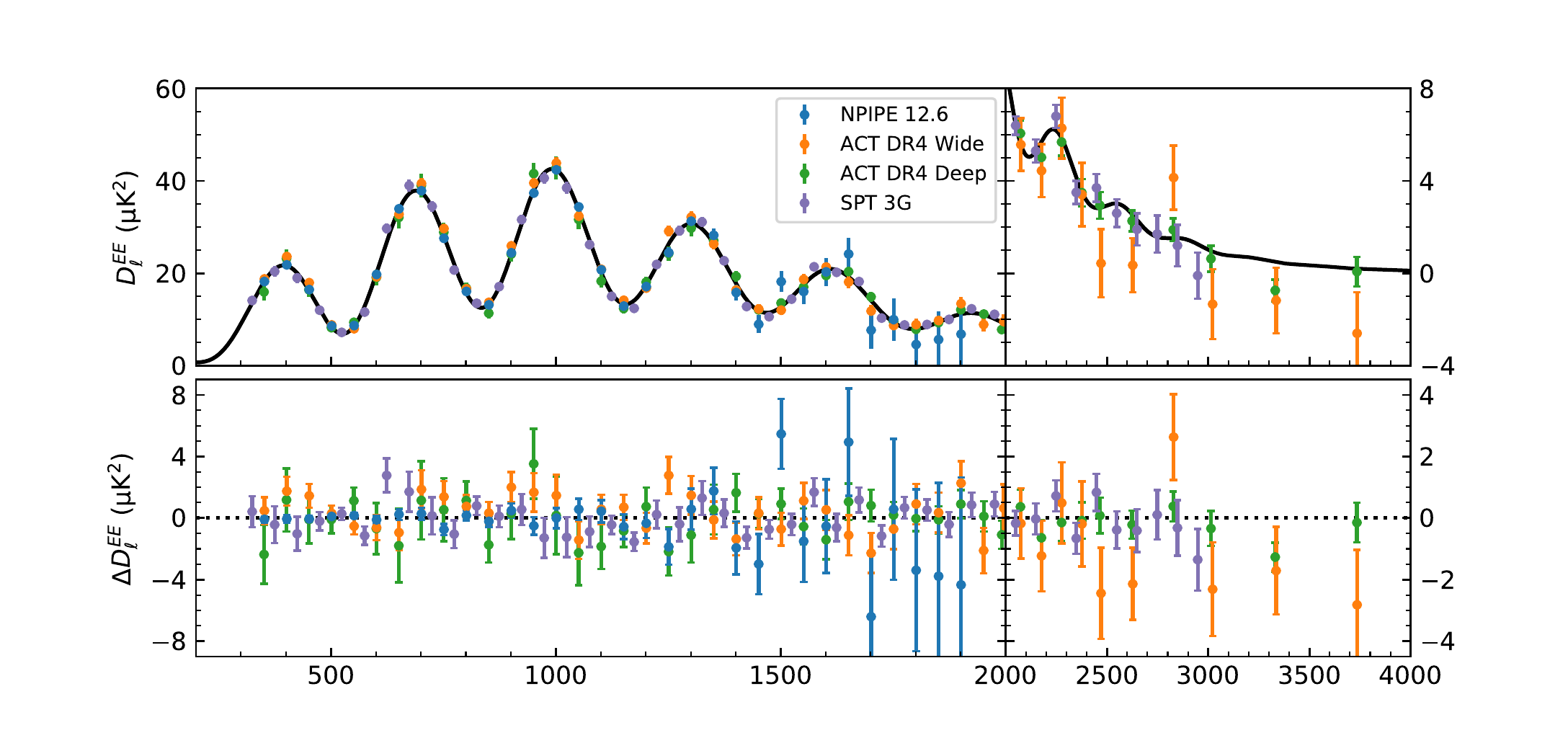}
  \caption{As Fig. \ref{fig:actsptTE} for EE.}
  \label{fig:actsptEE}
\end{figure*}
\subsection{\hillipop{}}
A limited analysis of the \npipe{} maps using the \hillipop{}\footnote{\url{https://github.com/planck-npipe/hillipop}} likelihood \citep{couchot17} and focusing on the tensor-to-scalar ratio $r$ has been published in \citet{tristram21}. \hillipop{} is also a pseudo-$C_\ell$ likelihood and uses \npipe{} data products. However the likelihood differs from ours in choice of masks, spectra, and foreground mitigation approach. Instead of cleaning spectra, \hillipop{} attempts to fit parametric models to uncleaned temperature and polarization spectra, including those using 100\GHz{} data. \citet{tristram21} find high-$\ell$ TT to give essentially identical results to PR3; full high-$\ell$ TE and EE results are not given there due to that paper's focus on $r$. We use the public \hillipop{} likelihood\footnote{Note that all references to \hillipop{} in this paper use \npipe{} data products.} with all default prior ranges on nuisance and foreground parameters together with our own priors on cosmological parameters to compare to our \camspec{} likelihood. Note that we continue to use \texttt{Commander} and \texttt{SimAll} at low multipoles. These results are shown alongside \plik{} and our baseline \camspec{} and \npipe{} constraints in Fig.\ \ref{fig:npipe-rd12-plik}. We find that \hillipop{} favors an even lower $\thetastar$ than PR4\_12.6; this is driven by polarization as shown in Fig.\ \ref{fig:hillipop-thetastar}, while the TT-only constraints (not shown) are similar to those of PR4\_12.6. Otherwise the \hillipop{} constraints do generally agree with our findings at the sigma level.

\subsection{ACT DR4 \& SPT-3G}
We present a simple comparison of our results to the latest ground-based data, ACT DR4 and SPT-3G. More detailed comparisons to \texttt{plik} are presented in the ACT \citep{choi20, aiola20} and SPT \citep{dutcher21} papers, as well as in \citet{handley20},  with a comparison to \camspec{} in the appendix of \eg{}. 

 For ACT each region, array, and season is independently calibrated to \planck{} PR3 temperature by cross-correlation. There is no separate polarization calibration; instead there is a free effective polarization efficiency in the likelihood. We compare to co-added CMB-only spectra with foregrounds and temperature-to-polarization leakage corrections marginalized out.
SPT also uses \planck{} PR3 cross-spectra to calibrate temperature of each sub-field before co-addition. They calibrate the final co-added spectra again to both TT and EE, with TE fixed. Foregrounds are not removed from the polarization spectra, but are reported to be negligibly small \citep{dutcher21}.

We show power spectra and residuals comparing our \npipe{} likelihood PR4\_12.6 to ACT DR4 and SPT-3G in Figs.\ \ref{fig:actTT}, \ref{fig:actsptTE}, and \ref{fig:actsptEE}. Table \ref{tab:actspt-chi2} presents $\chi^2$ values for the ACT and SPT data using the PR3\_12.6 and PR4\_12.6 best-fit $\lcdm$ models\footnote{$\chi^2$ values are calculated using the public ACT likelihood (\url{https://github.com/ACTCollaboration/pyactlike}) and a \texttt{python} port (\url{https://github.com/xgarrido/spt\_likelihoods}) of the public SPT-3G likelihood.}.
In TT we find generally good agreement between ACT and \planck{} but see that ACT Wide is consistently high by about 10-20$\muK^2$ in $D_\ell$ from $\ell$ of about 1000 to 1700. Both ACT Deep and Wide are also a bit low relative to \planck{} at very high multipoles (above 3500). These differences are reflected in somewhat high $\chi^2$ values in TT, slightly worse with the \npipe{} best-fit than with that of PR3. We interpret this as due to the calibration of ACT DR4 to the PR3 spectra. Indeed, allowing a free calibration parameter decreases the reduced $\chi^2$ in TT to 1.30 (PR4) and 1.28 (PR3), confirming that calibration explains most of the difference. 

In TE, like TT, the \npipe{} spectra are so similar to those from PR3 that previous comparisons to ground-based experiments generally still apply. By eye ACT TE appears to agree well with \npipe{}. However we note that \citet{choi20} found significant disagreement between PR3 and a subset of ACT TE spectra. There are no obvious features causing this disagreement, and further work is needed to identify its cause.
On the SPT side, there is a run of low points at $\ell < 700$, while higher multipoles appear to agree well.
In EE ACT DR4 Wide is consistently about 5$\muKsq$ low for $\ell > 2500$; otherwise all of the spectra agree well. The TEEE $\chi^2$ is very good, being within one sigma for both ACT and SPT. Overall we find, as for PR3, that the three experiments are broadly consistent but a more careful analysis with detailed map-based relative calibration would be necessary for a more quantitative comparison.
\renewcommand{\tabcolsep}{2pt}
\begin{table}
  \begin{tabular}{lllll}
    \hline
    Data         & PR4\_12.6 & PR3\_12.6 & ACT DR4 & SPT-3G TEEE\\
    \hline
    ACT TTTEEE   & 1.21 (2.43) & 1.16 (1.87) & 1.08 (0.88) & 1.22 (2.51) \\
    ACT TT       & 1.57 (3.62) & 1.48 (3.01) & 1.18 (1.15) & 1.37 (2.33) \\
    ACT TEEE     & 1.10 (0.92) & 1.09 (0.82) & 0.99 (-0.08) & 1.19 (1.79) \\
    SPT TEEE   & 0.98 (-0.34) & 0.98 (-0.32) & 1.02 (0.29) & 0.97 (-0.53) \\
    \hline
  \end{tabular}
  \caption{Reduced chi-squared $\rchisq$ and uncertainties for ACT and SPT data (rows) with $\lcdm$ best-fits from \planck{}, ACT, and SPT. The spectra used for each model match the data in all cases except the SPT best-fits, which are always TEEE. Multipole ranges are as in the likelihood for each experiment.}
  \label{tab:actspt-chi2}
\end{table}
\renewcommand{\tabcolsep}{6pt}

\section{Conclusions}
\label{sec:conclusions}

In this work we have presented new high-$\ell$ CMB temperature and polarization power spectra and likelihoods using the \planck{} PR4 `\npipe{}' maps. We find excellent consistency between the PR4 and PR3 power spectra, which translates to very good agreement on cosmological parameters as well. The most significant difference is in EE, for which \npipe{} prefers a lower value of $\thetastar$; however we find that the PR3\_12.6 best-fit $\lcdm$ model is still a good fit to the \npipe{} data.

The lower noise of the \npipe{} maps leads to tighter parameter constraints, with a \textasciitilde 10\% improvement in most $\lcdm$ parameters in TTTEEE due primarily to improvements in polarization. For $\lcdm$ extensions we find that, relative to PR3, \npipe{} polarization shrinks the error bars on $\omk$ and $A_L$ from EE by 40\% and 25\% respectively, and by 15\% and 8\% in TTTEEE. That these smaller error bars are accompanied by shifts toward the $\lcdm{}$ values continues the trend observed in \eg{} of decreasing the $\omk$ and $A_L$ tensions as more data is used, as would be expected if these pulls were due to a statistical fluctuation.
Overall, we conclude that \npipe{}, despite substantial differences in the mapmaking, is completely consistent with PR3 at the power spectrum and parameter levels, offering yet more evidence of the robustness of \planck{} data and the cosmological results inferred from \planck{}.
\section*{Acknowledgements}
We are grateful to the \npipe{} team and the wider \planck{} Collaboration for the data products used in this work. We thank Reijo Keskitalo for helpful correspondence and thank Anthony Challinor and John Peacock for useful comments on earlier drafts of this paper. Thanks also to the referee for helpful suggestions on the submitted manuscript.

ER acknowledges support from an Isaac Newton Studentship. SG acknowledges the award of a Kavli Institute Fellowship at KICC.

We gratefully acknowledge use of the following software packages: \textsc{pspipe} (\url{https://github.com/simonsobs/PSpipe}), \textsc{pspy} (\url{https://github.com/simonsobs/pspy}), \textsc{camb} \citep{Lewis:1999bs, Howlett:2012mh}, \textsc{cobaya} \citep{Torrado:2020xyz,BOBYQA,2018arXiv180400154C,2018arxiv181211343C}, \textsc{getdist} \citep{Lewis:2019xzd}, the MCMC sampler of \citet{Lewis:2002ah, Lewis:2013hha} using the fast-dragging procedure of \citet{Neal:2005}, \textsc{scipy} \citep{2020SciPy-NMeth}, \textsc{numpy} \citep{harris2020array}, \textsc{astropy} \citep{astropy:2013, astropy:2018}, \textsc{matplotlib} \citep{Hunter:2007}, \textsc{healpy} \citep{Zonca2019}, and \textsc{healpix} \citep{2005ApJ...622..759G}.

This work was performed in part using resources provided by the Cambridge Service for Data Driven Discovery (CSD3) operated by the University of Cambridge Research Computing Service (\url{www.csd3.cam.ac.uk}), provided by Dell EMC and Intel using Tier-2 funding from the Engineering and Physical Sciences Research Council (capital grant EP/T022159/1), and DiRAC funding from the Science and Technology Facilities Council (\url{www.dirac.ac.uk}).
This research also used resources of the National Energy Research Scientific Computing Center (NERSC), a U.S. Department of Energy Office of Science User Facility located at Lawrence Berkeley National Laboratory, operated under Contract No. DE-AC02-05CH11231.

For the purpose of open access, the authors have applied a Creative Commons Attribution (CC BY) licence to any Author Accepted Manuscript version arising.
This is a pre-copyedited, author-produced PDF of an article accepted for publication in the Monthly Notices of the Royal Astronomical Society following peer review. The version of record (MNRAS 517 (2022), 4620-4636) is available online at \url{https://academic.oup.com/mnras/advance-article-abstract/doi/10.1093/mnras/stac2744/6717656}.

\section*{Data Availability}
\label{sec:dataavailability}
The \planck{} and \npipe{} data products underlying the likelihoods presented here are publicly available at the PLA (\url{http://pla.esac.esa.int/pla/}) and at NERSC. The PR3\_12.6 and PR4\_12.6 likelihoods are both publicly available as internal likelihoods to \texttt{Cobaya} (\url{https://github.com/CobayaSampler/cobaya}), named \texttt{planck\_2018\_highl\_CamSpec2021} (PR3) and \texttt{planck\_NPIPE\_highl\_CamSpec} (PR4). 
\bibliographystyle{mnras}
\bibliography{Planck_bib,paper_draft,cobaya_bib}{}

\appendix
\section{Supplementary tables and figures}
\label{sec:appendix}
\begin{figure*}
  \centering
  \includegraphics[width=0.75\textwidth]{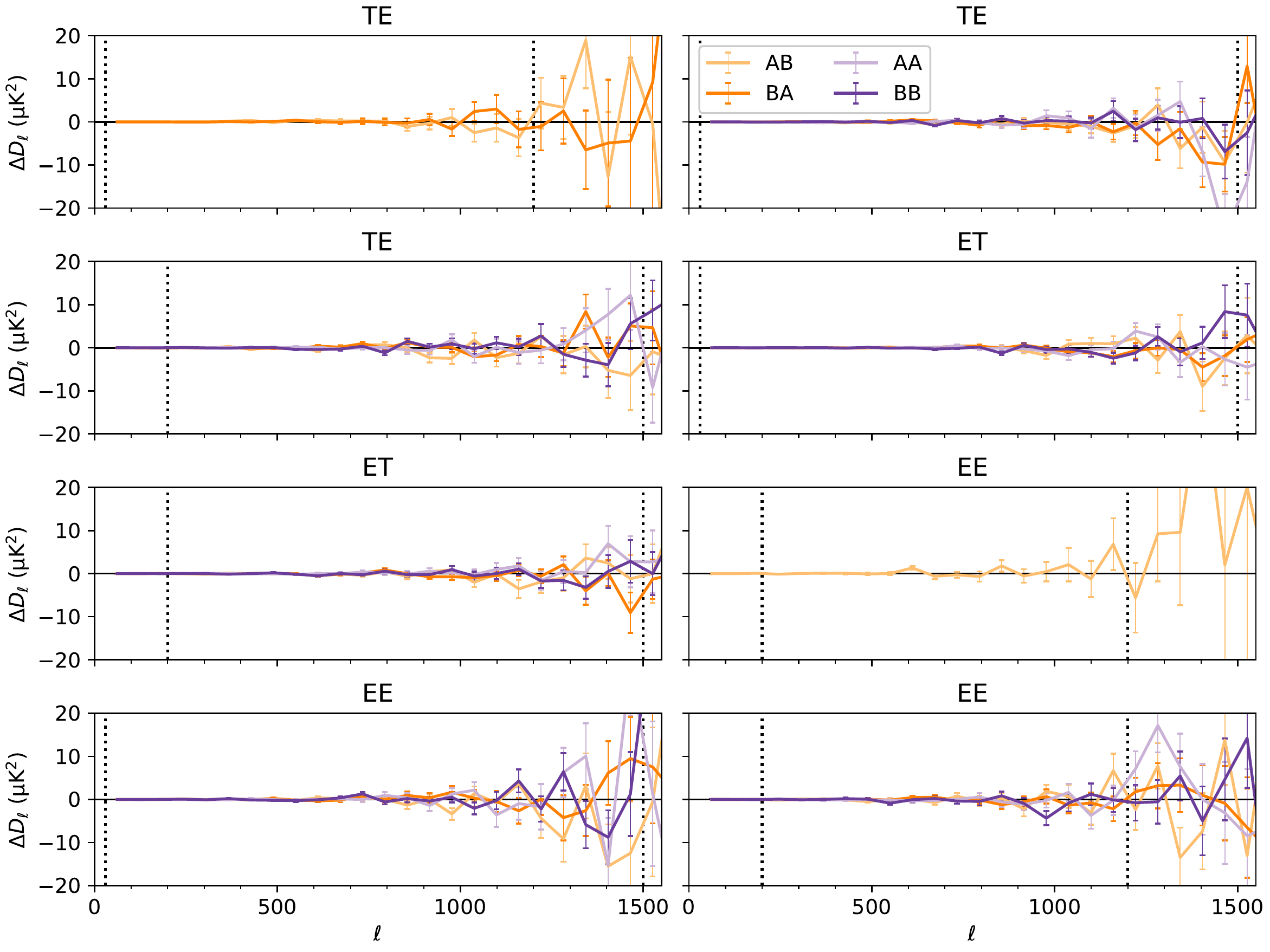}
  \includegraphics[width=0.75\textwidth]{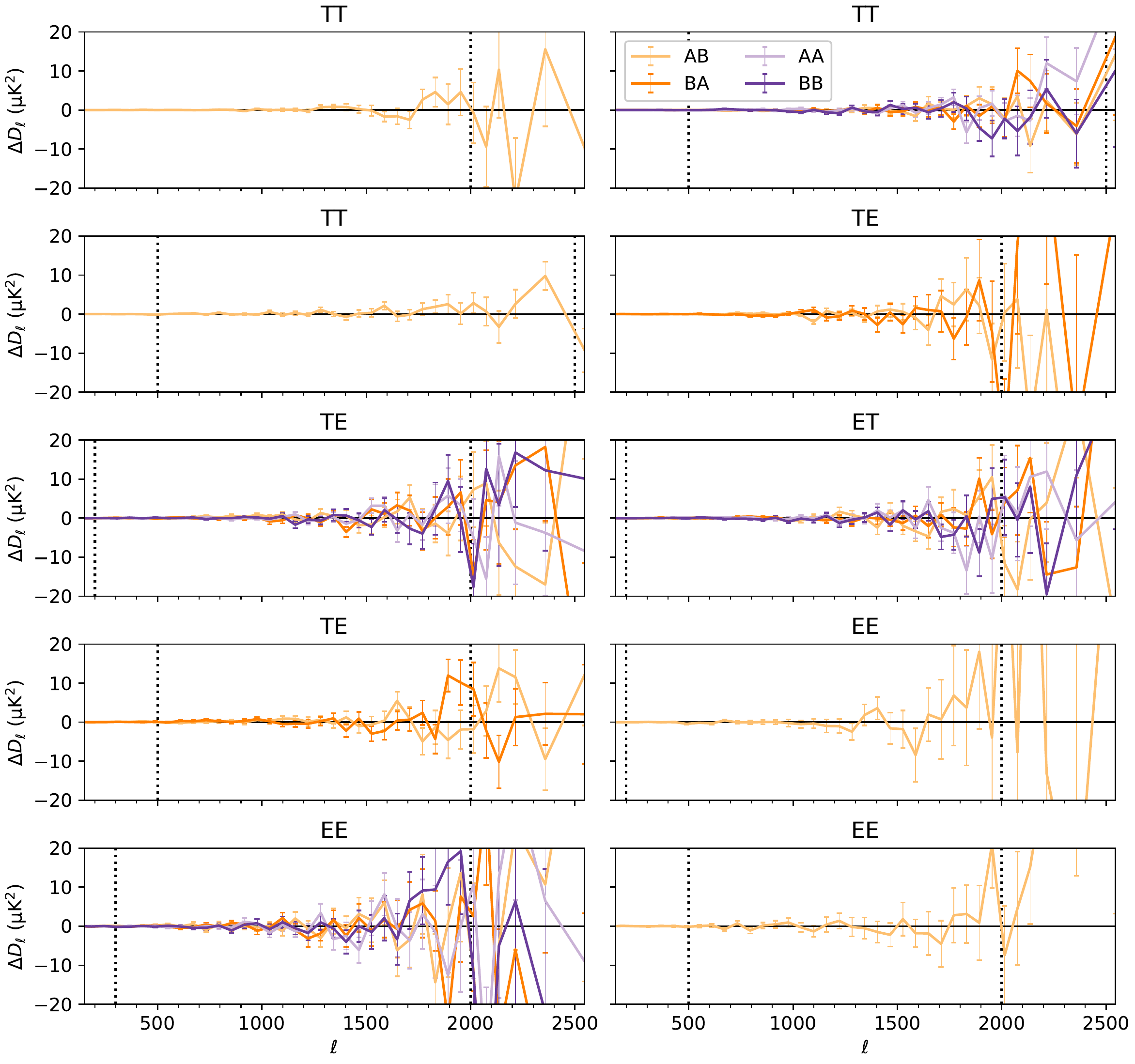}
  \caption{\npipe{} half-ring difference cross-spectra. Error bars show the standard deviation $\sigma / \sqrt{\Delta \ell}$ within each bandpower of bin-width $\Delta \ell$.}
  \label{fig:npipecorrnoisecross}
\end{figure*}

This appendix contains the following supplementary information: polarization spectrum calibration factors (Table \ref{tab:calibrations}), description of the multipole ranges used in the \camspec{} likelihoods (Table \ref{tab:multipole}), half-ring difference cross-spectra for \npipe{} (Fig.\ \ref{fig:npipecorrnoisecross}), and $\lcdm$ parameter constraints for PR4\_12.6 and PR3\_12.6 TT, TE, and EE likelihoods (Table \ref{tab:allparams}).
\begin{table}
  \centering
  \begin{tabular}{ccc}
    \hline Spectrum & TT & $c_k$ \\
    $143 \times 217$ & & $0.9992 \pm 0.0004$\\
    $217 \times 217$ & & $0.9988 \pm 0.0005$\\
    \hline Spectrum & EE index & $c_k$ \\
    $100\mathrm{A} \times 100\mathrm{B}$ & (1,2) & $0.988 \pm 0.007$ \\
    $100\mathrm{A} \times 143\mathrm{B}$ & (1,4) & $0.987 \pm 0.006$ \\
    $100\mathrm{A} \times 217\mathrm{B}$ & (1,6) & $0.970 \pm 0.007$ \\
    $100\mathrm{B} \times 143\mathrm{A}$ & (2,3) & $1.008 \pm 0.006$ \\
    $100\mathrm{B} \times 217\mathrm{A}$ & (2,5) & $0.955 \pm 0.007$ \\
    $143\mathrm{A} \times 143\mathrm{B}$ & (3,4) & $1.016 \pm 0.005$ \\
    $143\mathrm{A} \times 217\mathrm{B}$ & (3,6) & $0.975 \pm 0.006$ \\
    $143\mathrm{B} \times 217\mathrm{A}$ & (4,5) & $0.969 \pm 0.006$ \\
    $217\mathrm{A} \times 217\mathrm{B}$ & (5,6) & $0.941 \pm 0.007$ \\
    \hline Spectrum & TE index & $c_k$ \\
    $100\mathrm{A} \times 100\mathrm{B}$ & (1,2) & $0.985 \pm 0.009$ \\
    $100\mathrm{A} \times 143\mathrm{B}$ & (1,4) & $1.002 \pm 0.008$ \\
    $100\mathrm{A} \times 217\mathrm{B}$ & (1,6) & $0.985 \pm 0.009$ \\
    $100\mathrm{B} \times 143\mathrm{A}$ & (2,3) & $1.010 \pm 0.008$ \\
    $100\mathrm{B} \times 217\mathrm{A}$ & (2,5) & $0.967 \pm 0.009$ \\
    $143\mathrm{A} \times 143\mathrm{B}$ & (3,4) & $0.997 \pm 0.008$ \\
    $143\mathrm{A} \times 217\mathrm{B}$ & (3,6) & $0.980 \pm 0.009$ \\
    $143\mathrm{B} \times 217\mathrm{A}$ & (4,5) & $0.966 \pm 0.009$ \\
    $217\mathrm{A} \times 217\mathrm{B}$ & (5,6) & $0.985 \pm 0.009$ \\
    $100\mathrm{B} \times 100\mathrm{A}$ & (2,1) & $0.980 \pm 0.010$ \\
    $143\mathrm{B} \times 100\mathrm{A}$ & (4,1) & $0.984 \pm 0.010$ \\
    $217\mathrm{B} \times 100\mathrm{A}$ & (6,1) & $0.985 \pm 0.010$ \\
    $143\mathrm{A} \times 100\mathrm{B}$ & (3,2) & $0.983 \pm 0.009$ \\
    $217\mathrm{A} \times 100\mathrm{B}$ & (5,2) & $0.985 \pm 0.009$ \\
    $143\mathrm{B} \times 143\mathrm{A}$ & (4,3) & $1.009 \pm 0.008$ \\
    $217\mathrm{B} \times 143\mathrm{A}$ & (6,3) & $1.015 \pm 0.008$ \\
    $217\mathrm{A} \times 143\mathrm{B}$ & (5,4) & $0.998 \pm 0.008$ \\
    $217\mathrm{B} \times 217\mathrm{A}$ & (6,5) & $0.972 \pm 0.009$ \\
    \hline

  \end{tabular}
  \caption{Calibration factors measured from the \npipe{} spectra. Each co-added TT spectrum is multiplied by the given factor upon incorporation into the likelihood. Each TE and EE spectrum is divided (not multiplied) by the relevant factor before coaddition into the single TE and EE spectra of the likelihoods. Indices 1-6 correspond to 100A, 100B, 143A, 143B, 217A, and 217B respectively.}
  \label{tab:calibrations}
\end{table}

\begin{table}
\centering
\begin{tabular}{c|rrr}
  & \multicolumn{1}{c}{TT} & \multicolumn{1}{c}{TE} & \multicolumn{1}{c}{EE} \\
\hline
    $100 \times 100$ & \multicolumn{1}{c}{-}& $30-1200$  & $200-1200$ \\
    $100 \times 143$ & \multicolumn{1}{c}{-}& $30-1500$  & $30-1500$  \\
    $100 \times 217$ & \multicolumn{1}{c}{-}& $200-1500$ & $200-1200$ \\
    $143 \times 143$ & $30-2000$ & $30-2000$ & $200-2000$ \\
    $143 \times 217$ & $500-2500$ & $200-2000$ & $300-2000$ \\
    $217 \times 217$ & $500-2500$ & $500-2000$ & $500-2000$\\
    Total & $30-2500$ & $30-2000$ & $30-2000$\\
\end{tabular}
\caption{Multipole ranges used in the likelihoods described in this paper.}
\label{tab:multipole}
\end{table}

\renewcommand{\tabcolsep}{2pt}
\begin{table*}
  \begin{tabular}{ccccccc}
    \hline
& PR4\_12.6& PR4\_12.6& PR4\_12.6& PR3\_12.6& PR3\_12.6& PR3\_12.6\\
  & TT& TE& EE& TT& TE& EE\\
  \hline
$\Omega_\mathrm{b} h^2$                 &$0.02209\pm 0.00019$                    &$0.02223\pm 0.00017$                    &$0.02269\pm 0.00073$                    &$0.02215\pm 0.00020$                    &$0.02221\pm 0.00019$                    &$0.02316\pm 0.00093$                    \\
$\Omega_\mathrm{c} h^2$                 &$0.1196\pm 0.0018$                      &$0.1186\pm 0.0014$                      &$0.1200\pm 0.0031$                      &$0.1193\pm 0.0019$                      &$0.1191\pm 0.0016$                      &$0.1161\pm 0.0038$                      \\
$100\,\theta_\mathrm{MC}$               &$1.04083\pm 0.00039$                    &$1.04117\pm 0.00034$                    &$1.03934\pm 0.00055$                    &$1.04073\pm 0.00041$                    &$1.04142\pm 0.00040$                    &$1.04114\pm 0.00071$                    \\
$\tau$                                  &$0.0515\pm 0.0073$                      &$0.0490^{+0.0071}_{-0.0064}$            &$0.0491\pm 0.0071$                      &$0.0523\pm 0.0075$                      &$0.0490\pm 0.0071$                      &$0.0525\pm 0.0074$                      \\
$\ln(10^{10} A_\mathrm{s})$             &$3.034\pm 0.015$                        &$3.029\pm 0.019$                        &$3.044\pm 0.019$                        &$3.036\pm 0.015$                        &$3.025\pm 0.019$                        &$3.040\pm 0.020$                        \\
$n_\mathrm{s}$                          &$0.9626\pm 0.0052$                      &$0.9606\pm 0.0078$                      &$0.9619\pm 0.0086$                      &$0.9654\pm 0.0054$                      &$0.9581\pm 0.0089$                      &$0.985\pm 0.012$                        \\
\hline
$H_0$                                   &$67.26\pm 0.79$                         &$67.82\pm 0.62$                         &$67.1\pm 1.7$                           &$67.38\pm 0.83$                         &$67.71\pm 0.72$                         &$69.5\pm 2.2$                           \\
$100\,\thetastar$                       &$1.04107\pm 0.00038$                    &$1.04139\pm 0.00034$                    &$1.03951\pm 0.00054$                    &$1.04096\pm 0.00040$                    &$1.04164\pm 0.00040$                    &$1.04126\pm 0.00069$                    \\
$\Omega_\Lambda$                        &$0.685\pm 0.011$                        &$0.6921\pm 0.0085$                      &$0.681^{+0.023}_{-0.020}$               &$0.687\pm 0.011$                        &$0.6901\pm 0.0098$                      &$0.709^{+0.026}_{-0.021}$               \\
$\Omega_{\mathrm{m}}$                   &$0.315\pm 0.011$                        &$0.3078\pm 0.0085$                      &$0.319^{+0.020}_{-0.023}$               &$0.313\pm 0.011$                        &$0.3099\pm 0.0098$                      &$0.291^{+0.021}_{-0.026}$               \\
$\Omega_{\mathrm{m}} h^2$               &$0.1423\pm 0.0017$                      &$0.1415\pm 0.0014$                      &$0.1433\pm 0.0025$                      &$0.1421\pm 0.0018$                      &$0.1420\pm 0.0016$                      &$0.1399\pm 0.0031$                      \\
$\sigma_8$                              &$0.8060\pm 0.0082$                      &$0.8001\pm 0.0093$                      &$0.807\pm 0.012$                        &$0.8064\pm 0.0083$                      &$0.7997\pm 0.0096$                      &$0.798\pm 0.015$                        \\
$\sigma_8 \Omega_{\mathrm{m}}^{0.5}$    &$0.452\pm 0.011$                        &$0.4438\pm 0.0096$                      &$0.455\pm 0.021$                        &$0.451\pm 0.012$                        &$0.445\pm 0.011$                        &$0.430^{+0.024}_{-0.026}$               \\
$\sigma_8 \Omega_{\mathrm{m}}^{0.25}$   &$0.604\pm 0.010$                        &$0.5959\pm 0.0095$                      &$0.606\pm 0.018$                        &$0.603\pm 0.011$                        &$0.597\pm 0.010$                        &$0.586\pm 0.022$                        \\
$z_\mathrm{re}$                         &$7.42\pm 0.76$                          &$7.12^{+0.78}_{-0.63}$                  &$7.05^{+0.75}_{-0.63}$                  &$7.49\pm 0.76$                          &$7.13^{+0.78}_{-0.65}$                  &$7.25\pm 0.72$                          \\
$10^9 A_{\mathrm{s}}$                   &$2.079\pm 0.032$                        &$2.069\pm 0.038$                        &$2.099\pm 0.039$                        &$2.083\pm 0.032$                        &$2.061\pm 0.038$                        &$2.092\pm 0.042$                        \\
$10^9 A_{\mathrm{s}} e^{-2\tau}$        &$1.875\pm 0.013$                        &$1.876\pm 0.022$                        &$1.902\pm 0.024$                        &$1.876\pm 0.013$                        &$1.868\pm 0.023$                        &$1.883\pm 0.027$                        \\
$\mathrm{Age}/\mathrm{Gyr}$             &$13.826\pm 0.031$                       &$13.798\pm 0.026$                       &$13.812\pm 0.091$                       &$13.822\pm 0.033$                       &$13.794\pm 0.029$                       &$13.69\pm 0.12$                         \\
$r_{\mathrm{drag}}$                     &$147.52\pm 0.43$                        &$147.62\pm 0.36$                        &$146.76\pm 0.47$                        &$147.54\pm 0.43$                        &$147.50\pm 0.40$                        &$147.28\pm 0.67$                        \\
\hline
$A_\mathrm{Planck}$                        &$1.0004\pm 0.0024$                      &$0.99998\pm 0.0025$                     &$1.0000\pm 0.0024$                      &$1.0003\pm 0.0024$                      &$1.0000\pm 0.0025$                      &$1.0000\pm 0.0024$                      \\
$c_\mathrm{TE}$                                &-                                       &$0.9997\pm 0.0092$                      &-                                       &-                                       &$0.9997\pm 0.0091$                      &-                                       \\
$c_\mathrm{EE}$                                &-                                       &-                                       &$1.0000\pm 0.0091$                      &-                                       &-                                       &$1.0004\pm 0.0091$                      \\
$A^{\rm power}_{143}$                   &$19.1^{+2.0}_{-2.8}$                    &-                                       &-                                       &$18.4^{+2.2}_{-2.9}$                    &-                                       &-                                       \\
$A^{\rm power}_{143\times217}$          &$9.8^{+1.9}_{-2.7}$                     &-                                       &-                                       &$9.2^{+2.1}_{-2.9}$                     &-                                       &-                                       \\
$A^{\rm power}_{217}$                   &$12.9^{+1.9}_{-2.7}$                    &-                                       &-                                       &$14.0^{+2.1}_{-2.9}$                    &-                                       &-                                       \\
$\gamma^{\rm power}_{143}$              &$0.97^{+0.23}_{-0.17}$                  &-                                       &-                                       &$1.05^{+0.31}_{-0.25}$                  &-                                       &-                                       \\
$\gamma^{\rm power}_{143\times217}$     &$1.54^{+0.53}_{-0.61}$                  &-                                       &-                                       &$1.32^{+0.50}_{-0.72}$                  &-                                       &-                                       \\
$\gamma^{\rm power}_{217}$              &$1.39\pm 0.42$                          &-                                       &-                                       &$1.27\pm 0.41$                          &-                                       &-                                       \\
\hline
\end{tabular}
\caption{PR4\_12.6 and PR3\_12.6 parameter constraints in $\lcdm$. We report mean values and 68\% confidence intervals. TTTEEE is shown in Table \ref{tab:npipeparams}.}
\label{tab:allparams}
\end{table*}
\renewcommand{\tabcolsep}{6pt}
\bsp
\label{lastpage}
\end{document}